\documentclass[10pt, journal, final]{Components/MetaFiles/IEEEtran}

\usepackage{gensymb}
\usepackage{dsfont}
\usepackage{amsmath,amssymb,amsfonts,amsthm}
\usepackage{epsfig}
\usepackage{cite}
\usepackage{hhline}
\usepackage{multirow}
\usepackage{xcolor}
\usepackage{makecell}
\usepackage{subcaption}
\usepackage{capt-of}
\usepackage[labelformat=simple]{subcaption}

\usepackage{enumerate}
\usepackage{enumitem}
\usepackage{color}
\usepackage{graphicx}
\usepackage{algorithmic}
\usepackage[ruled]{algorithm}
\usepackage{booktabs}
\usepackage{bm}
\usepackage{tikz}
\usepackage[nolist,printonlyused]{acronym} 
\usepackage{colortbl}

\usepackage[T1]{fontenc}
\usepackage[utf8]{inputenc}
\usepackage{pgfplots}
\usepackage{comment} 

\pgfplotsset{compat=newest}
\usetikzlibrary{plotmarks}
\usepgfplotslibrary{patchplots}

\usetikzlibrary{automata,arrows,positioning,calc,matrix,decorations.markings,decorations.pathreplacing}

\makeatletter
\newcommand{\vast}{\bBigg@{3}}
\newcommand{\Vast}{\bBigg@{4}}
\makeatother
\renewcommand{\IEEEQED}{\IEEEQEDopen}

\showoutput
\begin{document}

\title{Pilot Precoding and Combining in \\ Multiuser MIMO Networks}
\author{Nima N. Moghadam,           \textit{Student Member, IEEE},
        Hossein Shokri-Ghadikolaei, \textit{Student Member, IEEE},\\
        Gabor Fodor,                \textit{Senior Member, IEEE},
        Mats Bengtsson,             \textit{Senior Member, IEEE}, and
        Carlo Fischione,            \textit{Member, IEEE}
}


\newtheorem{theorem}{Theorem}
\newtheorem{defin}{Definition}
\newtheorem{prop}{Proposition}
\newtheorem{lemma}{Lemma}
\newtheorem{corollary}{Corollary}
\newtheorem{alg}{Algorithm}
\newtheorem{remark}{Remark}
\newtheorem{result}{Result}
\newtheorem{conjecture}{Conjecture}
\newtheorem{example}{Example}
\newtheorem{notations}{Notations}
\newtheorem{assumption}{Assumption}

\newcommand{\be}{\begin{equation}}
\newcommand{\ee}{\end{equation}}
\newcommand{\ba}{\begin{array}}
\newcommand{\ea}{\end{array}}
\newcommand{\bea}{\begin{eqnarray}}
\newcommand{\eea}{\end{eqnarray}}
\newcommand{\combin}[2]{\ensuremath{ \left( \ba{c} #1 \\ #2 \ea \right) }}
\newcommand{\diag}{{\mathrm{diag}}}
\newcommand{\rank}{{\mbox{rank}}}
\newcommand{\dom}{{\mathrm{dom{\color{white!100!black}.}}}}
\newcommand{\range}{{\mbox{range{\color{white!100!black}.}}}}
\newcommand{\image}{{\mbox{image{\color{white!100!black}.}}}}
\newcommand{\herm}{^{\mbox{\scriptsize H}}}  
\newcommand{\sherm}{^{\mbox{\tiny H}}}       
\newcommand{\tran}{^{\mathrm{\scriptsize T}}}  
\newcommand{\tranIn}{^{\mbox{-\scriptsize T}}}  
\newcommand{\card}{{\mbox{\textbf{card}}}}
\newcommand{\asign}{{\mbox{$\colon\hspace{-2mm}=\hspace{1mm}$}}}
\newcommand{\ssum}[1]{\mathop{ \textstyle{\sum}}_{#1}}

\newcommand{\vbar}{\raisebox{.17ex}{\rule{.04em}{1.35ex}}}
\newcommand{\vbarind}{\raisebox{.01ex}{\rule{.04em}{1.1ex}}}
\newcommand{\D}{\ifmmode {\rm I}\hspace{-.2em}{\rm D} \else ${\rm I}\hspace{-.2em}{\rm D}$ \fi}
\newcommand{\T}{\ifmmode {\rm I}\hspace{-.2em}{\rm T} \else ${\rm I}\hspace{-.2em}{\rm T}$ \fi}
\newcommand{\B}{\ifmmode {\rm I}\hspace{-.2em}{\rm B} \else \mbox{${\rm I}\hspace{-.2em}{\rm B}$} \fi}
\newcommand{\Hil}{\ifmmode {\rm I}\hspace{-.2em}{\rm H} \else \mbox{${\rm I}\hspace{-.2em}{\rm H}$} \fi}
\newcommand{\C}{\ifmmode \hspace{.2em}\vbar\hspace{-.31em}{\rm C} \else \mbox{$\hspace{.2em}\vbar\hspace{-.31em}{\rm C}$} \fi}
\newcommand{\Cind}{\ifmmode \hspace{.2em}\vbarind\hspace{-.25em}{\rm C} \else \mbox{$\hspace{.2em}\vbarind\hspace{-.25em}{\rm C}$} \fi}
\newcommand{\Q}{\ifmmode \hspace{.2em}\vbar\hspace{-.31em}{\rm Q} \else \mbox{$\hspace{.2em}\vbar\hspace{-.31em}{\rm Q}$} \fi}
\newcommand{\Z}{\ifmmode {\rm Z}\hspace{-.28em}{\rm Z} \else ${\rm Z}\hspace{-.38em}{\rm Z}$ \fi}

\newcommand{\sgn}{\mathrm {sgn}}
\newcommand{\var}{\mathrm {var}}
\newcommand{\E}{\mathbb{E}}
\newcommand{\cov}{\mathrm {cov}}
\renewcommand{\Re}{\mathrm {Re}}
\renewcommand{\Im}{\mathrm {Im}}
\newcommand{\cum}{\mathrm {cum}}

\renewcommand{\vec}[1]{{\mathbf{#1}}}     
\newcommand{\vecsc}[1]{\mbox {\boldmath \scriptsize $#1$}}     
\newcommand{\itvec}[1]{\mbox {\boldmath $#1$}}
\newcommand{\itvecsc}[1]{\mbox {\boldmath $\scriptstyle #1$}}
\newcommand{\gvec}[1]{\mbox{\boldmath $#1$}}

\newcommand{\balpha}{\mbox {\boldmath $\alpha$}}
\newcommand{\bbeta}{\mbox {\boldmath $\beta$}}
\newcommand{\bgamma}{\mbox {\boldmath $\gamma$}}
\newcommand{\bdelta}{\mbox {\boldmath $\delta$}}
\newcommand{\bepsilon}{\mbox {\boldmath $\epsilon$}}
\newcommand{\bvarepsilon}{\mbox {\boldmath $\varepsilon$}}
\newcommand{\bzeta}{\mbox {\boldmath $\zeta$}}
\newcommand{\boldeta}{\mbox {\boldmath $\eta$}}
\newcommand{\btheta}{\mbox {\boldmath $\theta$}}
\newcommand{\bvartheta}{\mbox {\boldmath $\vartheta$}}
\newcommand{\biota}{\mbox {\boldmath $\iota$}}
\newcommand{\blambda}{\mbox {\boldmath $\lambda$}}
\newcommand{\bmu}{\mbox {\boldmath $\mu$}}
\newcommand{\bnu}{\mbox {\boldmath $\nu$}}
\newcommand{\bxi}{\mbox {\boldmath $\xi$}}
\newcommand{\bpi}{\mbox {\boldmath $\pi$}}
\newcommand{\bvarpi}{\mbox {\boldmath $\varpi$}}
\newcommand{\brho}{\mbox {\boldmath $\rho$}}
\newcommand{\bvarrho}{\mbox {\boldmath $\varrho$}}
\newcommand{\bsigma}{\mbox {\boldmath $\sigma$}}
\newcommand{\bvarsigma}{\mbox {\boldmath $\varsigma$}}
\newcommand{\btau}{\mbox {\boldmath $\tau$}}
\newcommand{\bupsilon}{\mbox {\boldmath $\upsilon$}}
\newcommand{\bphi}{\mbox {\boldmath $\phi$}}
\newcommand{\bvarphi}{\mbox {\boldmath $\varphi$}}
\newcommand{\bchi}{\mbox {\boldmath $\chi$}}
\newcommand{\bpsi}{\mbox {\boldmath $\psi$}}
\newcommand{\bomega}{\mbox {\boldmath $\omega$}}

\newcommand{\bolda}{\mbox {\boldmath $a$}}
\newcommand{\bb}{\mbox {\boldmath $b$}}
\newcommand{\bc}{\mbox {\boldmath $c$}}
\newcommand{\bd}{\mbox {\boldmath $d$}}
\newcommand{\bolde}{\mbox {\boldmath $e$}}
\newcommand{\boldf}{\mbox {\boldmath $f$}}
\newcommand{\bg}{\mbox {\boldmath $g$}}
\newcommand{\bh}{\mbox {\boldmath $h$}}
\newcommand{\bp}{\mbox {\boldmath $p$}}
\newcommand{\bq}{\mbox {\boldmath $q$}}
\newcommand{\br}{\mbox {\boldmath $r$}}
\newcommand{\bt}{\mbox {\boldmath $t$}}
\newcommand{\bu}{\mbox {\boldmath $u$}}
\newcommand{\bv}{\mbox {\boldmath $v$}}
\newcommand{\bw}{\mbox {\boldmath $w$}}
\newcommand{\bx}{\mbox {\boldmath $x$}}
\newcommand{\by}{\mbox {\boldmath $y$}}
\newcommand{\bz}{\mbox {\boldmath $z$}}

\newenvironment{Ex}
{\begin{adjustwidth}{0.04\linewidth}{0cm}
\begingroup\small
\vspace{-1.0em}
\raisebox{-.2em}{\rule{\linewidth}{0.3pt}}
\begin{example}
}
{
\end{example}
\vspace{-5mm}
\rule{\linewidth}{0.3pt}
\endgroup
\end{adjustwidth}}

\newcommand{\Hossein}[1]{{\textcolor{magenta}{\emph{**Hossein: #1**}}}}
\newcommand{\Nima}[1]{{\textcolor{magenta}{\emph{**Nima: #1**}}}}
\newcommand{\Comment}[1]{{\textcolor{blue}{\emph{**Comment: #1**}}}}
\newcommand{\Challenge}[1]{{\textcolor{red}{#1}}}
\newcommand{\NEW}[1]{{\textcolor{blue}{#1}}}


\makeatletter
\let\save@mathaccent\mathaccent
\newcommand*\if@single[3]{%
  \setbox0\hbox{${\mathaccent"0362{#1}}^H$}%
  \setbox2\hbox{${\mathaccent"0362{\kern0pt#1}}^H$}%
  \ifdim\ht0=\ht2 #3\else #2\fi
  }
\newcommand*\rel@kern[1]{\kern#1\dimexpr\macc@kerna}
\newcommand*\widebar[1]{\@ifnextchar^{{\wide@bar{#1}{0}}}{\wide@bar{#1}{1}}}
\newcommand*\wide@bar[2]{\if@single{#1}{\wide@bar@{#1}{#2}{1}}{\wide@bar@{#1}{#2}{2}}}
\newcommand*\wide@bar@[3]{%
  \begingroup
  \def\mathaccent##1##2{%
    \let\mathaccent\save@mathaccent
    \if#32 \let\macc@nucleus\first@char \fi
    \setbox\z@\hbox{$\macc@style{\macc@nucleus}_{}$}%
    \setbox\tw@\hbox{$\macc@style{\macc@nucleus}{}_{}$}%
    \dimen@\wd\tw@
    \advance\dimen@-\wd\z@
    \divide\dimen@ 3
    \@tempdima\wd\tw@
    \advance\@tempdima-\scriptspace
    \divide\@tempdima 10
    \advance\dimen@-\@tempdima
    \ifdim\dimen@>\z@ \dimen@0pt\fi
    \rel@kern{0.6}\kern-\dimen@
    \if#31
      \overline{\rel@kern{-0.6}\kern\dimen@\macc@nucleus\rel@kern{0.4}\kern\dimen@}%
      \advance\dimen@0.4\dimexpr\macc@kerna
      \let\final@kern#2%
      \ifdim\dimen@<\z@ \let\final@kern1\fi
      \if\final@kern1 \kern-\dimen@\fi
    \else
      \overline{\rel@kern{-0.6}\kern\dimen@#1}%
    \fi
  }%
  \macc@depth\@ne
  \let\math@bgroup\@empty \let\math@egroup\macc@set@skewchar
  \mathsurround\z@ \frozen@everymath{\mathgroup\macc@group\relax}%
  \macc@set@skewchar\relax
  \let\mathaccentV\macc@nested@a
  \if#31
    \macc@nested@a\relax111{#1}%
  \else
    \def\gobble@till@marker##1\endmarker{}%
    \futurelet\first@char\gobble@till@marker#1\endmarker
    \ifcat\noexpand\first@char A\else
      \def\first@char{}%
    \fi
    \macc@nested@a\relax111{\first@char}%
  \fi
  \endgroup
}
\makeatother

\newcommand{\tr}{{\rm{tr}}}  
\newcommand{\vect}{{\rm{vec}}}  
\newcommand{\cond}{\kappa}  
\newcommand{\Nb}{M}
\newcommand{\Nu}{N}
\newcommand{\Np}{L}
\newcommand{\I}{\mathbf{I}}
\newcommand{\Hh}{\widehat{\mathbf{H}}}
\newcommand{\Ht}{\widetilde{\mathbf{H}}}
\newcommand{\Gh}{\widehat{\mathbf{G}}}
\newcommand{\Gt}{\widetilde{\mathbf{G}}}
\newcommand{\Ab}{\mathbf{B}}
\newcommand{\Au}{\mathbf{U}}
\newcommand{\Rb}{\mathbf{R}^\text{B}}
\newcommand{\Ru}{\mathbf{R}^\text{U}}
\newcommand{\setU}{\mathcal{U}}
\newcommand{\setK}{\mathcal{K}}
\newcommand{\PilotEnergy}{\rho_\tau}
\newcommand{\DataEnergy}{\rho_d}
\newcommand{\NoisePower}{\sigma_z^2}
\newcommand{\CG}{\sigma^2} 
\newcommand{\Tt}{T_\tau} 
\newcommand{\Td}{T_d} 
\newcommand{\Tc}{T_c} 
\newcommand{\sx}{\mathtt{x}} 
\newcommand{\so}{\text{nPuC}} 
\newcommand{\st}{\text{PuC}} 
\newcommand{\sth}{\text{PC}} 
\newcommand{\Pt}{\breve{\mathbf{P}}}
\newcommand{\Wt}{\breve{\mathbf{W}}}
\newcommand{\Rt}{\widetilde{\mathbf{R}}}

\newcommand{\Po}{\mathbf{P}^{s_1}}
\newcommand{\Pth}{\mathbf{P}^{s_3}}

\def\REV#1{\textcolor{red}{#1}} 
\newcommand{\GF}[1]{\textcolor{red}{#1}} 
\newcommand{\note}[1]{\textcolor{red}{#1}} 

\graphicspath{{Components/Figures/}}

\begin{acronym}[LTE-Advanced]
  \acro{2G}{Second Generation}
  \acro{3G}{3$^\text{rd}$~Generation}
  \acro{3GPP}{3$^\text{rd}$~Generation Partnership Project}
  \acro{4G}{4$^\text{th}$~Generation}
  \acro{5G}{5$^\text{th}$~Generation}
  \acro{AoA}{angle of arrival}
  \acro{AoD}{angle of departure}
  \acro{BER}{bit error rate}
  \acro{BF}{beamforming}
  \acro{BLER}{BLock Error Rate}
  \acro{BPC}{Binary Power Control}
  \acro{BPSK}{Binary Phase-Shift Keying}
  \acro{BRA}{Balanced Random Allocation}
  \acro{BS}{base station}
  \acro{CAP}{Combinatorial Allocation Problem}
  \acro{CAPEX}{Capital Expenditure}
  \acro{CBF}{Coordinated Beamforming}
  \acro{CS}{Coordinated Scheduling}
  \acro{CSI}{channel state information}
  \acro{CSIT}{channel state information at the transmitter}
  \acro{CSIR}{channel state information at the receiver}
  \acro{D2D}{device-to-device}
  \acro{DCA}{Dynamic Channel Allocation}
  \acro{DE}{Differential Evolution}
  \acro{DFT}{Discrete Fourier Transform}
  \acro{ULA}{uniform linear array}
  \acro{DIST}{Distance}
  \acro{DL}{downlink}
  \acro{DMA}{Double Moving Average}
  \acro{DMRS}{Demodulation Reference Signal}
  \acro{D2DM}{D2D Mode}
  \acro{DMS}{D2D Mode Selection}
  \acro{DPC}{Dirty Paper Coding}
  \acro{DRA}{Dynamic Resource Assignment}
  \acro{DSA}{Dynamic Spectrum Access}
  \acro{FDD}{frequency division duplexing}
  \acro{LTE}{Long Term Evolution}
  \acro{LSAS}{large scale antenna system}
  \acro{LTE}{Long Term Evolution}
  \acro{METIS}{Mobile Enablers for the Twenty-Twenty Information Society}
  \acro{MIMO}{Multiple-input multiple-output}
  \acro{MMSE}{minimum mean squared error}
  \acro{MU-MIMO}{multiuser MIMO}
  \acro{SUMIMO}{single-user multiple-input multiple-output}
  \acro{MISO}{multiple-input single-output}
  \acro{mmWave}{millimeter-wave}
  \acro{MRC}{maximum ratio combining}
  \acro{MS}{mode selection}
  \acro{MSE}{mean square error}
  \acro{MTC}{machine type communications}
  \acro{NSPS}{national security and public safety}
  \acro{NWC}{network coding}
  \acro{PC}{pilot contamination}
  \acro{PHY}{physical layer}
  \acro{QoS}{Quality of Service}
  \acro{QPSK}{Quadri-Phase Shift Keying}
  \acro{RAISES}{Reallocation-based Assignment for Improved Spectral Efficiency and Satisfaction}
  \acro{RAN}{Radio Access Network}
  \acro{RA}{Resource Allocation}
  \acro{RAT}{Radio Access Technology}
  \acro{RB}{resource block}
  \acro{RF}{radio frequency}
  \acro{SINR}{signal-to-noise-and-interference ratio}
  \acro{SNR}{signal-to-noise ratio}
  \acro{STC}{space-time coding}
  \acro{TDD}{time division duplexing}
  \acro{UE}{user equipment}
  \acro{UL}{uplink}
  \acro{VUE}{vehicular user equipment}
  \acro{V2X}{vehicle-to-vehicle and vehicle-to-infrastructure}
  \acro{ZF}{Zero-Forcing}
  \acro{ZMCSCG}{Zero Mean Circularly Symmetric Complex Gaussian}
\end{acronym}

\begin{acronym}
	\acro{PSD}{positive semi-definite}
	\acro{SVD}{singular value decomposition}
	\acro{EVD}{eigen value decomposition}
\end{acronym} 

\maketitle
\begin{abstract}
Although the benefits of precoding and combining data signals are widely recognized,
the potential of these techniques for pilot transmission
is not fully understood.
This is particularly relevant for multiuser multiple-input multiple-output (MU-MIMO) cellular systems using millimeter-wave (mmWave) communications, where multiple antennas have to be used both at the transmitter and the receiver to overcome the severe path loss.
In this paper, we characterize the gains of pilot precoding and combining in terms of
channel estimation quality and achievable data rate.
Specifically, we consider three uplink pilot transmission scenarios in
a mmWave MU-MIMO cellular system:
1)~non-precoded and uncombined, 2)~precoded but uncombined, and 3)~precoded and combined.
We show that a simple precoder
that utilizes only the second-order statistics of the channel reduces the variance of the channel estimation error
by a factor that is proportional to the number of \ac{UE} antennas.
We also show that using a linear combiner designed based on the second-order statistics of the channel
significantly reduces multiuser interference and provides the possibility of reusing some pilots.
Specifically, 
in the large antenna regime,
pilot precoding and combining help to accommodate
a large number of \acp{UE} in one cell, significantly improve channel estimation quality,
boost the signal-to-noise ratio of the UEs located close to the cell edges,
alleviate pilot contamination, and address the imbalanced coverage of pilot and data signals.
\end{abstract}


\section{Introduction}\label{sec: introductions}
\ac{MIMO} systems that employ a large
number of antenna ports
at wireless access points
are a rapidly maturing technology.
The \ac{3GPP} is currently studying the details of technology enablers
and performance benefits of deploying large scale antenna systems 
that support up to 64 antenna ports
at cellular \acp{BS} \cite{36.897}.
Moreover, higher frequency bands, such as \ac{mmWave},
will naturally employ
large-scale antenna systems~\cite{Han:15}.
With \ac{mmWave} antennas, the physical array size can be greatly
reduced due to the decrease in wavelength.
Therefore, it is expected that wireless systems employing even greater number of
antenna ports will be deployed in \ac{mmWave} bands, making massive \ac{MIMO} systems
a practical reality.

In addition to the usage of many antenna ports at the \ac{BS},
also user equipments (UEs) compliant with existing and emerging wireless standards
are also employing a growing number of receive and transmit antennas.
For example, current
\acp{UE} of the \ac{3GPP} Long Term Evolution systems can employ up to four
antennas
for transmit and receive diversity as well as
for spatial multiplexing \cite{36.101}.
Clients of the IEEE 802.11ac standard can employ up to eight antenna elements~\cite{bejarano2013ieee}.
For 5G systems, we expect high-end \acp{UE} supporting high order of modulation and coding schemes
and a greater number of receive and transmit antennas~\cite{METIS-Book}.
Moreover, due to the high path loss at the mmWave frequencies, exploiting multiple antennas in addition to
the spatial precoding and combining is considered to have an essential role for establishing and maintaining a robust communication link~\cite{Kutty2016}.
Nonetheless, most of the significant investigations
in massive \ac{MU-MIMO} assume that
the 
\acp{BS} or 
access points
serve a lower number of \emph{single-antenna} \acp{UE}.
In those studies, such an assumption is considered non-restrictive because the spatial
precoding of the user data streams 
boosts the achieved \ac{SINR}~\cite{Rusek:13},~\cite{Marzetta:10}.
While transmit precoding for the downlink transmission is the key to achieve high spectral efficiency,
precoding in the uplink direction
has not been considered in
these works.
This is true not only for uplink data transmission, but also for
the transmission of uplink pilot signals that are used to acquire
both \ac{CSIT} and \ac{CSIR} at the \ac{BS}.
A direct consequence of the lack of precoding of pilot signals is
that the majority of the existing schemes consider one orthogonal pilot sequence per transmit antenna,
which gives
several systematic problems -- illustrated in the sequel --
and may limit the efficiency and future use cases of
\ac{MU-MIMO} systems, especially in \ac{mmWave} networks.


Acquiring accurate \ac{CSI}, either at the transmitter or at the receiver, is among the main bottlenecks of
massive \ac{MIMO} systems and faces three main challenges:
\emph{i)} scalability of the number of pilots, \emph{ii)} performance at low \ac{SNR}, and \emph{iii)} pilot contamination~\cite{Bjornson:16},~\cite{Choi:14}.
The length of training sequences for channel estimation,
in the traditional ``one orthogonal pilot sequence per transmit antenna'' scheme
scales up (at least) linearly with the number of transmit antennas~\cite{HassibiHochwald2003}.
Assuming that the number of BS antennas is larger than the combined number of UE antennas,
channel estimation in uplink imposes shorter training sequences.
However, this scheme requires the principle of channel reciprocity to hold, which is valid
only for \ac{TDD} mode and when the duplexing time is much shorter
than the coherence time of the channel.
Thus, realizing massive MIMO systems in \ac{FDD} mode is a well-known challenge~\cite{Bjornson:16}.
Even in \ac{TDD} mode, this scheme may not be feasible when a massive number of multiple-antenna UEs are present,
which is an important use case of \ac{mmWave} networks.
In such a case, the entire coherence budget may be used only for the channel estimation procedure,
depriving the data transmission phase from valuable coherent time and frequency resources.
Thus, a pilot transmission scheme that is scalable
with the number of transmit antennas is especially desirable in mmWave systems.
Moreover, due to the central role of the uplink pilot signals in the \ac{CSI} acquisition, the constrained UE power and the lack of uplink precoding gains
may limit the performance of mmWave systems with large antenna arrays.
This leads to an \emph{imbalanced coverage of pilot and data signals}
(a.k.a imbalance between uplink and downlink coverage~\cite{LTEBook},~\cite{Bhaskara:14}), 
where the range at which
reasonable data rates can be maintained differs from
the one at which pilot signals can be detected.
This problem is particularly important in \ac{mmWave} networks due to severe channel attenuations~\cite{shokri2015mmWavecellular}.
Therefore, a good pilot transmission scheme should address the imbalance in pilot-data coverage.

The above-mentioned technical challenges, which limit the achievable data rate of massive MIMO systems,
are exacerbated by pilot contamination,
defined as the interference in the pilot signals~\cite{Rusek:13},~\cite{Marzetta:10},~\cite{Elijah:15}.
Several methods to alleviate pilot contamination have been proposed
and demonstrated~\cite{Bjornson:16,Saxena:15,Liu:15,Yin:13}.
In the context of multi-cell networks,
the results of~\cite{Yin:13} suggest that channel covariance-aware pilot assignment to \acp{UE} can completely remove the pilot contamination effects in the limit of large number of \ac{BS} antennas.
The intuition is that if a selected UE exhibits multipath angles of arrival (AoA) at its serving \ac{BS},
which do not overlap with the AoAs of UEs in the neighboring cells, these UEs can reuse the same pilot sequence as the selected \ac{UE}, without any contamination among the pilots of different cells.
In~\cite{Liu:15}, the UEs within each cell employ the same pilot sequence while the UEs in the different cells are assigned orthogonal pilot sequences.
The impact of intra-cell pilot contamination is mitigated later in the downlink data transmission phase.
However, in this case extra precoding matrices should be designed at the BSs, which add to the complexity of data transmission, specially when the number of antennas grows large.
The other studies have not investigated the pilot reuse within one cell, despite that such a reuse, together with employing multiple antenna elements at the \ac{UE}, has the potential to substantially improve the spectral efficiency.


In this paper, we argue that all the aforementioned problems
-- scalability, poor performance at low \ac{SNR}, and pilot contamination --
can be substantially alleviated by employing multiple antennas at the~\acp{UE},
together with pilot precoding
and combining.
We show the benefits of pilot precoding and combining for general massive MU-MIMO systems and for \ac{mmWave} systems in particular, where these benefits are substantial.
Our investigation is motivated not only by the ongoing standards development and the need to boost the uplink \ac{SNR},
but also by the expectation that pilot precoding may achieve better spatial separation of \acp{UE} served in the same and surrounding cells.
Specifically, we focus on the uplink of a MU-\ac{MIMO} system
and consider three pilot transmission scenarios: 1) non-precoded and uncombined ($\so$), which serves as a baseline scenario, 2) precoded but uncombined ($\st$), and 3) precoded and combined ($\sth$). We use these three scenarios to study the gains of pilot precoding and combining in terms of channel estimation quality and the achievable data rate.
Our extensive mathematical and numerical analysis results in the following key findings:
\begin{itemize}
\item
Pilot precoding in $\st$ substantially improves the channel estimation quality that can be achieved by the baseline $\so$ scenario.
In particular, we show that the channel estimation error variance
can be reduced by a factor of $\delta = \Nu/\Np$ for large values of $\Nu$,
where $\Nu$ is the number of \ac{UE} antenna elements,
and $\Np$ is the rank of the channel between each UE and the BS.
Hence, pilot precoding leads to a large improvement in the channel estimation performance
of systems with a large number of antennas
and low rank channels (which is the case in \ac{mmWave} networks)
where $\delta$ is large.
\item A simple combiner of the pilot sequence, in the $\sth$ scenario
that uses only the second-order statistics of the channel substantially reduces intra-cell multiuser interference.
In the systems with a large number of antennas at the BS, such as in mmWave networks, this interference can be canceled completely.
Consequently in $\sth$, pilots can be reused within one cell, as opposed to orthogonal pilot sequences used in the $\so$ and $\st$ scenarios.
This potential translates into the possibility of channel estimation using shorter training sequences (or equivalently serving more \acp{UE} without extra training sequences), which in turn enhances the network spectral efficiency.
\item Unlike the baseline scenario, $\so$, the number of pilot symbols needed in $\st$ and $\sth$, is not dependent on the number of antenna elements at the transmitters.
This benefit enables the realization of massive \ac{MIMO} systems in both \ac{FDD} and \ac{TDD} deployments.
\item
In $\sth$, when the number of antennas at the BS and UEs goes to infinity,
which may resemble a wireless backhauling scenario,
we conclude the following asymptotic results:
\begin{enumerate}
\item
the effects of pilot contamination vanish;
\item
a multi-cell network can be modeled by multiple uncoordinated single-cell systems
with no performance loss (that is, without a performance penalty due to the lack of coordination); and
\item
channel estimation in the entire network can be done with only one pilot symbol.
We also characterize the above gains in the finite antenna regime.
\end{enumerate}
\end{itemize}

Our investigations can
contribute to answering
the following fundamental questions
related to large scale antenna systems in general and \ac{mmWave} networks in particular~\cite{Daniels:10}:
How can we improve the quality of channel estimation for a fixed (finite) number of \ac{BS} antennas?
Can we mitigate the effects of pilot contamination and can we reduce the need for multicell coordination?
Can we utilize massive antenna arrays in \ac{FDD} systems? How many orthogonal pilot sequences
do we need for a given number of -- possibly multiple antenna -- \acp{UE}?
The rest of this paper is organized as follows. Section~\ref{Sec:Model} describes the system model, including the channel and the signal model.
Section~\ref{sec: Channel Estimation} analyzes three distinct channel estimation techniques
that differ in terms of complexity and achievable channel estimation quality.
Section~\ref{sec:Data Transmission} studies the data transmission phase that makes use of the acquired \ac{CSI} using the schemes discussed in Section~\ref{sec: Channel Estimation}.
Section~\ref{sec: FurtherDiscussions}
presents additional engineering insights, and
Section~\ref{Sec:Conc} concludes the paper.
Useful definitions and lemmas, which are used throughout the paper, are given in Appendix~A.

\emph{Notations:} Capital bold letters denote matrices and lower bold letters denote vectors.
The superscript $[\vec{X}]^*$, $[\vec{X}]\tran$, $[\vec{X}]\herm$ and $[\vec{X}]^\dagger$
stand for the conjugate, transpose, transpose conjugate and Moore-Penrose pseudoinverse of $\vec{X}$, respectively.
The subscript $[\vec{X}]_{i,j}$ denotes entry of $\vec{X}$ at row $i$ and column~$j$.
$[\vec{X}]_{:,i}$ represents column $i$ of $\vec{X}$.
$\vec{I}$ is the identity matrix with the appropriate size,
$\vec{I}_{x}$ is the identity matrix with size $x$,
$\vect(\vec{X})$ is
the vectorization of matrix $\vec{X}$, and $\diag(\vec{x})$ is a diagonal matrix
with entries $\vec{x}$. The Hadamard product (element-wise product),
Kronecker product and Khatri-Rao product of matrices $\mathbf{X}$
and $\mathbf{Y}$ are denoted by $\mathbf{X}\circ\mathbf{Y}$,
$\mathbf{X}\otimes\mathbf{Y}$, and $\mathbf{X}\odot \mathbf{Y}$ respectively.
Table~\ref{table: notations} lists the main symbols used throughout this paper.
\begin{table}[t]
  \centering
  \caption{Summary of main notations. }\label{table: notations}
{
\renewcommand{\arraystretch}{1.4}
  {
   \begin{tabular}{|@{}c@{}|l|}
\hline
   \textbf{Symbol} & \textbf{Definition} \\ \hline
    $\vec{H}_k \in\mathbb{C}^{\Nb \times \Nu}$ & Uplink channel matrix between UE $k$ and the BS\\ \hline
    $\Au_k \in\mathbb{C}^{\Nu \times \Np}$ & Antenna response of UE $k$ toward its AoDs \\ \hline
    $\Ab_k \in\mathbb{C}^{\Nb \times \Np}$ & Antenna response of the BS toward its AoAs \\ \hline
    $\vec{G}_k \in\mathbb{C}^{\Np \times \Np}$ & Complex path gains between UE $k$ and the BS  \\ \hline
    $\widehat{\vec{X}}, \widetilde{\vec{X}}$ & An estimation of $\vec{X}$, and corresponding estimation error \\ \hline
    $\cond(\vec{X})$ & Condition number of $\vec{X}$ \\ \hline
	$~\Nu, \Nb~$ & Number of antennas at UEs and at the BS\\ \hline
    $\Np$ & Number of paths between every UE and the BS \\ \hline
    $K$ & Number of UEs in the cell\\ \hline
    $\Tc$ & Coherence time of fast fading \\ \hline
    $T_s$ & Coherence time of slow fading \\ \hline
    $\Tt$ & Number of symbols transmitted during channel estimation \\ \hline
	$\Td$ & Number of symbols transmitted during data transmission \\ \hline
    ${\rho}_{\tau}, {\rho}_{d}$ & Total energy used for pilot/data transmission \\ \hline
    $1/\CG_k$ & Path-loss between UE $k$ and the \ac{BS}  \\ \hline
    $\NoisePower$ & Variance of a Gaussian noise \\ \hline
\end{tabular}}
}
\end{table}


\section{System Model}
\label{Sec:Model}
We consider the uplink of a single-cell multiuser MIMO network where a BS with $\Nb$ antennas is serving $K$ UEs each equiped with $\Nu$ antennas. We comment on the extension of our framework to a multi-cell network in Section~\ref{sec: FurtherDiscussions}.
Note that our system model is valid for both access and backhaul layers. In the backhaul scenario, the BS label could refer to
a common gateway and the UE label could refer to
small or macro cell BSs. Without loss of generality, in the following, we use the terminology of the access layer.


\subsection{Channel Model}
We assume a narrow-band block-fading channel between the BS and each UE, where the channels are relatively
constant for one fading block, with duration $T_c$ channel uses,
and they change to statistically independent values in the next block.
However,
we assume that the second-order statistics of the channel
remains unchanged for $T_s$ channel uses, where $T_s\gg T_c$.
Moreover, we consider a cluster channel model~\cite{el2014spatially} with $\Np$ paths between the BS and each UE. This model can be easily transformed into the well-known virtual channel model~\cite{sayeed2002deconstructing}.
Let $g^{i}_k$ be the complex gain of path $i$ between the BS and UE $k$, which includes both path-loss and small scale fading. In particular, $\{g^{i}_k\}$ for all $i\in\{1,\ldots,\Np\}$ are independent and identically distributed random variables drawn from distribution $\mathcal{CN}(0,\CG_k)$ where $1/\CG_k$ is the path-loss between the BS and UE $k$~\cite{Akdeniz2014MillimeterWave}.
It consists of a constant attenuation, a distance dependent attenuation, and a large scale log-normal fading.
The uplink channel matrix between the BS and UE $k$ is
\be \label{eq:geometry-based channel model}
\vec{H}_{k} = \sqrt{\frac{\Nb\Nu}{\Np}} \, \sum\limits_{i=1}^{\Np}{g^{i}_{k}\,{\vec{b}\left(\theta^{i}_{k}\right)} \vec{u}\herm\left(\phi^{i}_{k}\right) } = \mathbf{B}_k \mathbf{G}_k \mathbf{U}_k\herm \in \mathbb{C}^{\Nb\times\Nu} \:,
\ee
where $\theta^i_k$ and $\phi^i_k$ are the AoA an AoD of path $i$ between the BS and UE $k$, respectively.
Parameters $\vec{b} \in \mathbb{C}^{\Nb}$ and $\vec{u} \in \mathbb{C}^{\Nu}$ represent the normalized array response vectors of the BS's and UEs' antenna arrays, respectively, $\mathbf{B}_k = [\mathbf{b}(\theta_k^1),\dots,\mathbf{b}(\theta_k^{\Np})]$,
$\mathbf{U}_k = [\mathbf{u}(\phi_k^1),\dots,\mathbf{u}(\phi_k^{\Np})]$, and $\mathbf{G}_k  \in \mathbb{C}^{\Np \times \Np}$ is a diagonal matrix whose $i$-th diagonal entry is $g_{k}^{i}\sqrt{\Nb \Nu/\Np}$. The channel can be written in the vectorized format as \cite{brewer1978kronecker}
\be
\label{eq:channel model}
\vect\left(\mathbf{H}_k\right) = \left(\Au_k^{*}\otimes\Ab_k\right)\vect\left(\mathbf{G}_k\right) = \left(\Au_k^*\odot\Ab_k\right)\mathbf{g}_k \:,
\ee
where $\mathbf{g}_k$ is the principle diagonal of $\mathbf{G}_k$ and $\mathbf{A}_1\odot \mathbf{A}_2$ represents the Khatri-Rao product of $\mathbf{A}_1$, and $\mathbf{A}_2$ (see Definition~\ref{def: Hadamard-Kronecker-Khatri} in Appendix A).
For the sake of tractability in the asymptotic performance analysis,
we assume an antenna configuration at the BS and UEs which satisfies
\be\label{eq: antenna_config}
\lim_{\Nb\to\infty} \mathbf{b}(\theta)\herm\mathbf{b}(\phi) =
\lim_{\Nu\to\infty} \mathbf{u}(\theta)\herm\mathbf{u}(\phi) =
\begin{cases}
1& \theta = \phi\:,\\
0& \text{otherwise}\:,
\end{cases}
\ee
for any $\theta, \phi\in \mathbb{R}$. These conditions hold for uniform linear array antennas as well as randomly positioned antenna elements in the arrays. Although our general framework does not necessarily require~\eqref{eq: antenna_config} to hold, our asymptotic performance analysis
remains tractable if~\eqref{eq: antenna_config} holds.


When the AoA's and AoD's are given, channel $\mathbf{H}_k$ is zero-mean circularly symmetric Gaussian with
covariance matrix $\mathbf{R}_k$, which is defined based on the column stacking of the channel matrix.
Therefore, given $\theta^{i}_{k}$'s  and $\phi^{i}_{k}$'s, $\vect\left(\mathbf{H}_k\right)\sim\mathcal{CN}(\boldsymbol{0},\mathbf{R}_k)$.
Using \eqref{eq:channel model}, the covariance matrix of channel $\mathbf{H}_k$ can be found as
\begin{align}
\label{eq: R_H}
\mathbf{R}_{k} & = \left(\Au_k^{*}\odot\Ab_k\right)\E\left[\mathbf{g}_k \mathbf{g}_k\herm \right] \left(\Au_k^*\odot\Ab_k\right)\herm \nonumber \\
& = \delta \Nb \sigma_k^2  \left(\Au_k^*\odot\Ab_k\right) \left(\Au_k^*\odot\Ab_k\right)\herm \:,
\end{align}
where ${\delta = \Nu/\Np}$.
It can be seen from \eqref{eq: R_H} that the rank of $\mathbf{R}_k$ is the same as the rank of $\Au_k^*\odot\Ab_k$, which is equal to $\Np$.

\subsection{Signal Model}
Within one fading block, the baseband received signal vector at the BS is
\be \label{eq: baseband system model}
\mathbf{y}(t) = \sum_{k=1}^{K} \mathbf{H}_k \mathbf{x}_k(t) + \mathbf{z}(t) \;,
\ee
where $\mathbf{x}_k(t) \in \mathbb{C}^{\Nu}$ and $\mathbf{z}(t) \sim \mathcal{CN}(\boldsymbol{0},\NoisePower\I_{\Nb})$ represent the signal vector transmitted from UE $k$ and the receiver noise at the BS, respectively, at channel use $t$.

\begin{figure}
 \centering
 \includegraphics[width=0.9\columnwidth]{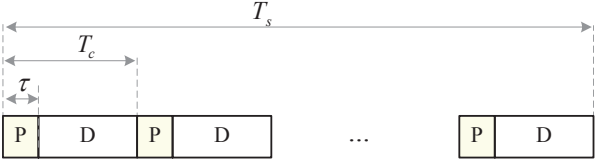}
 \caption{Superframe structure. ``P'' and ``D'' show pilot and data transmission phases.
 $T_c$ is the channel coherence time.
 $T_s$ is the superframe duration over which AoA and AoD remain unchanged,
 but the fast fading component of the channel changes per channel coherence time $\Tt$.}
 \label{fig: SuperframeStructure}
\end{figure}
As shown in Fig.~\ref{fig: SuperframeStructure}, the signal transmission within each fading block consists of two phases:
\emph{pilot transmission phase} with duration $T_{\tau}$ and \emph{data transmission phase} with duration $T_d = T_c-T_\tau$.
In the following, we will elaborate on these phases.
\subsubsection{Pilot Transmission}
To estimate the uplink channel, UE~$k$ transmits pilot matrix $\mathbf{P}_k$ over $\Tt$ channel uses.
Let $\PilotEnergy$ be the total energy devoted to the pilot transmission by each UE within each fading block, namely
\be \label{eq: pilot energy constraint}
\tr\left(\mathbf{P}_k \mathbf{P}_k\herm\right) = \PilotEnergy , \quad k=1,2,\ldots,K \:.
\ee
UE $k$ precodes its uplink pilots using spatial filter $\mathbf{V}_k$.
The received signal at the BS is then combined using spatial filter $\mathbf{W}_k$ in order to estimate the corresponding channel.
Collectively, the filtered received signal at the BS during the pilot transmission phase is
\be \label{eq: pilot transmission}
\mathbf{Y}_{\tau,k} = \mathbf{W}_k\herm\sum_{j=1}^{K}\mathbf{H}_j\mathbf{V}_j\mathbf{P}_j + \mathbf{W}_k\herm \mathbf{Z}_{\tau},
\ee
where $\mathbf{Z}_{\tau} \in \mathbb{C}^{\Nb\times \Tt}$ is the BS noise matrix with independent and identically distributed entries modeled as
$\vect(\mathbf{Z}_{\tau}) \sim \mathcal{CN}(\boldsymbol{0},\NoisePower\I_{\Nb \Tt})$.
Note that $\mathbf{P}_k$ and $\mathbf{Y}_{\tau,k}$ have $\Tt$ columns,
while the number of rows depends on the dimensions of the spatial filters $\mathbf{V}_k$ and $\mathbf{W}_k$, respectively.
In the next section, we will discuss pilot precoding and combining scenarios with different dimensions for the spatial filters.

\subsubsection{Data Transmission Phase}
In this phase of the signal transmission, all the \acp{UE}
transmit their precoded data symbols to the BS simultaneously.
The data precoding filters are designed at the BS using the estimated channels and are fed back to the UEs through the feedback links.
Each \ac{UE} transmits its data symbols in $\Np$ streams over $\Td$ channel uses with constant average power
$\DataEnergy/\Td$, where $\DataEnergy$ denotes the total energy devoted for data transmission by each UE within each fading block.
We assume that the data symbols are independent and identically distributed as zero-mean circularly symmetric complex Gaussian random variables.
Let $\mathbf{s}_k \in \mathbb{C}^{\Np}$ represent the data vector of UE $k$;
then $\mathbf{s}_k\sim \mathcal{CN}(\mathbf{0},{\DataEnergy}/{\Td}\I_{\Np})$.

The data vector $\mathbf{s}_k$ is precoded before transmission using the linear spatial filter $\mathbf{F}_k\in\mathbb{C}^{\Nu\times \Np}$.
The received signal at the BS during the data transmission phase is
\be \label{eq: data transmission}
\mathbf{y}_{d} = \sum_{k=1}^{K} \mathbf{H}_k \mathbf{F}_k\mathbf{s}_k + \mathbf{z}_d,
\ee
where $\mathbf{z}_d \sim\mathcal{CN}(\boldsymbol{0},\NoisePower\I_{\Nb})$ is the thermal noise at the BS.

\section{Channel Estimation}
\label{sec: Channel Estimation}
In this section, we study the channel estimation quality for three different pilot transmission scenarios:
\begin{itemize}
\item \emph{non-precoded and uncombined pilot transmission} ($\so$);
\item \emph{precoded and uncombined pilot transmission} ($\st$); and
\item \emph{precoded and combined pilot transmission} ($\sth$).
\end{itemize}
In the $\so$ baseline scenario, every antenna element of every \ac{UE} sends a unique pilot sequence
(typically orthogonal to other pilot sequences).
$\st$ allows precoding of the pilot signals at the transmitters (\acp{UE} in our case), but the receiver (\ac{BS}) will not jointly process the signals received by its $\Nb$ antenna elements.
Finally, $\sth$ permits both precoders and combiners in the pilot transmission phase. Using \eqref{eq: pilot transmission} and Lemma~\ref{lemma: vectorization} in Appendix~A, the vectorized received signal in pilot transmission scenario $\sx \in \{\so,\st,\sth\}$ is
\be \label{eq: vectorized received pilot}
\vect\left(\mathbf{Y}_{\tau,k}^{(\sx)}\right) = \sum_{j=1}^{K} \left(\Pt_{kj}^{(\sx)}\right)\herm \vect\left(\mathbf{H}_j\right) + \Wt_k \vect\left(\mathbf{Z}_{\tau}\right),
\ee
where $\big(\Pt^{(\sx)}_{kj}\big)\herm = \big(\mathbf{V}_j \mathbf{P}_j^{(\sx)}\big)\tran \otimes \mathbf{W}_k\herm$
and $\Wt_k = \I_{\Tt} \otimes \mathbf{W}_k\herm$.

In all the three scenarios, we use \ac{MMSE} estimates of the channel, assuming that the spatial filters are known at the BS.
Using \cite[Equation~(6)]{Bjornson2010} and noting that the mean values of our channel matrices are zero,
the MMSE estimate of channel $\mathbf{H}_k$ for scenario $\sx$,
can be expressed as
\be \label{eq: MMSE channel estimate}
\vect\!\left(\!\Hh_k^{(\sx)}\!\right)\!\!=\! \mathbf{R}_k \Pt_k^{(\sx)} \!\!
\left(\sum_{j=1}^{K}\!\left(\!\Pt_{kj}^{(\sx)}\!\right)^{\!\!\mbox{\scriptsize H}}
\mathbf{R}_j \Pt_{kj}^{(\sx)} \!+\! \NoisePower \Wt_{\!k}  \!\right)^{\!\!\!\!-1}\!\!\!\!\!
\vect\!\left(\!\mathbf{Y}_{\tau,k}^{(\sx)}\!\right),
\ee
where $\Pt^{(\sx)}_{k} = \Pt^{(\sx)}_{kk}$.

In the rest of this section, we drive the channel estimate in each pilot transmission scenario.
Based on these derivations, we will then analyze and compare
the channel estimation performance in the studied scenarios.

\subsection{Non-precoded and Uncombined Pilot Transmission ($\so$)}
In the first pilot transmission scenario, used as a benchmark,
orthogonal training sequences are transmitted from the different UEs' antenna elements\cite{Marzetta:10}.
Moreover, in this scenario, neither pilot precoding at the UEs nor pilot combining at the BS is performed.
On the positive side, this scenario requires no prior information about the channel.
On the negative side, this traditional scheme requires at least $T_\tau^{(\so)} = K \Nu$ resource elements for the pilot transmission phase,
which can grow large as the number of antenna elements per UE increases, e.g., in mmWave networks~\cite{Rangan2014Millimeter}.
Moreover, the lack of transmit/receive antenna gains reduces SNR of individual pilots, decreasing the channel estimation quality,
which negatively affects the data precoding performance of the BS and UEs.
Therefore, $\so$ does not meet any of the pilot transmission criteria that we mentioned earlier, i.e., scalability and balanced data-pilot coverage.
Considering the pilot energy constraint~\eqref{eq: pilot energy constraint}, we have
\be\label{eq: pilot matrix_s1}
\mathbf{P}_{k}^{(\so)}\left(\mathbf{P}_{j}^{(\so)}\right)\herm =
\begin{cases}
 \frac{\PilotEnergy}{\Nu} \I_{\Nu} & k = j\:, \\
 \boldsymbol{0} & \mbox{otherwise}\:,
\end{cases}
\ee
where $\mathbf{P}_{k}^{(\so)} \in \mathbb{C}^{\Nu\times K \Nu}$ is the pilot matrix transmitted from UE $k$ in scenario $\so$.
Substituting the pilot symbols of \eqref{eq: pilot matrix_s1} and replacing the spatial filters by identity matrices
in \eqref{eq: MMSE channel estimate}, the MMSE estimate of the vectorized channel is
\be
\begin{split}
\label{eq: MMSE channel estimate s1}
\vect\left(\Hh_k^{(\so)}\right) =&\mathbf{R}_{k}\Pt^{(\so)}_k
\left(\!\!\left(\!\Pt^{(\so)}_k\!\right)^{\!\!\mbox{\scriptsize H}} \!
\mathbf{R}_{k} \Pt^{(\so)}_k \!\!+\! \NoisePower \I\!\right)^{\!\!\!-1} \\
&\hspace{5mm}\cdot \vect\left(\mathbf{Y}_{\tau}^{(\so)}\right),
\end{split}
\ee
where $\Pt_k^{(\so)} = \big(\mathbf{P}_k^{(\so)}\big)^*\otimes\I_{\Nb}$. Note
that $\{\Pt_k^{(\so)}\}_{k=1}^{K}$ inherit the orthogonality property of
$\{\mathbf{P}_k^{(\so)}\}_{k=1}^{K}$,
and therefore the signals received from UE $j\neq k$ can be canceled out in the process of channel estimation for UE $k$.

Define the estimation error matrix of scenario $\so$ by ${\Ht^{(\so)}_{k}=\mathbf{H}_{k}-\Hh^{(\so)}_{k}}$.
Then, in the following proposition, we find its covariance under the traditional non-precoded uncombined pilot transmission scenario:
\begin{prop} \label{prop: covariance of error_s1}
Consider the system model of scenario $\so$. Suppose that the channel is estimated with orthogonal pilots given by~\eqref{eq: pilot matrix_s1} and an MMSE estimator given by~\eqref{eq: MMSE channel estimate s1}.
The covariance matrix of the channel estimation error is
\be
\label{eq: covariance of error_s1}
\begin{split}
\Rt_k^{(\so)} & =
\E\left[{\vect\left(\Ht^{(\so)}_k\right)} {\vect\left(\Ht^{(\so)}_k\right)}\herm\right] \\
& =\delta \Nb \CG_k \left(\Au_k^* \odot \Ab_k \right)  \\
&~~~\cdot \left(\I_{\Np} \!+\!  \Nb\zeta_k \left(\mathbf{R}_{\Au_k}\tran\!\circ\!\mathbf{R}_{\Ab_k}\right) \right)^{\!\!-1}\!\left(\Au_k^* \!\odot\! \Ab_k \right)\herm ,
\end{split}
\ee
where $\mathbf{R}_{\mathbf{U}_k} = \Au_k\herm\Au_k$, $\mathbf{R}_{\mathbf{B}_k} =
\Ab_k\herm\Ab_k$,
\be
\zeta_k = \frac{\PilotEnergy \CG_k}{\Np \NoisePower},
\ee
and the expectation is taken over the distribution of the random channel and the received noise.
\end{prop}
\begin{IEEEproof}
A proof is given in Appendix~B.
\end{IEEEproof}

The following proposition characterizes the normalized \ac{MSE} in scenario $\so$ as an indicator of the channel estimation quality.
\begin{prop}\label{prop: bounds-on-estimation-error-S1}
Consider the covariance of the MMSE channel estimation error in Proposition~\ref{prop: covariance of error_s1}.
The corresponding normalized MSE, $e_k^{(\so)} = \tr(\Rt_k^{(\so)})/\tr(\mathbf{R}_k)$, is bounded as
\be\label{eq: NMSE_s1}
\frac{1}{1+\Nb \zeta_k} \left[1 - \frac{\epsilon_k^{(\so)} }
{\Nb \zeta_k}\right]^{\!+} \!\!\le e_k^{(\so)} \!\le\!
\frac{1}{1+\Nb\zeta_k} \;,
\ee
where
\begin{equation*}
\epsilon_k^{(\so)} = \frac{\left(1-\cond\left(\I_{\Np} +\Nb\zeta_k\;\mathbf{R}_{\Au_k}\tran \circ \mathbf{R}_{\Ab_k} \right)\right)^2}
{4\cond\left(\I_{\Np} + \Nb\zeta_k\; \mathbf{R}_{\Au_k}\tran \circ \mathbf{R}_{\Ab_k} \right)} \ge 0 \:,
\end{equation*}
and $\left[a\right]^{+} = \max(a,0)$ for $a \in \mathbb{R}$.
\end{prop}

\begin{corollary}\label{corollary: NMSE_s1}
As $\Nu \to \infty$ (so $\mathbf{R}_{\Au_k} \to \I_{\Np}$),
both upper and lower bounds in~\eqref{eq: NMSE_s1} become tight in the sense that
\be
e_k^{(\so)} \to (1+\Nb\zeta_k)^{-1}.
\ee
\end{corollary}

\begin{figure}
\centering
\hspace{-0.95cm}
\begin{subfigure}[b]{0.39\columnwidth}
  {\footnotesize 
%
%
\definecolor{mycolor1}{rgb}{0.30588,0.39608,0.58039}%
\definecolor{mycolor2}{rgb}{0.85000,0.32500,0.09800}%
\definecolor{mycolor3}{rgb}{0.00000,0.74902,0.74902}%
\begin{tikzpicture}

\begin{axis}[%
width=1\textwidth,
height=1\textwidth,
at={(0\textwidth,0\textwidth)},
scale only axis,
xmode=log,
xmin=8,
xmax=1024,
ymin=-10,
ymax=-2,
xminorticks=true,
xlabel={$\Nu$},
ylabel shift={-4pt},
ylabel={Average normalized MSE (dB)},
axis background/.style={fill=white},
legend style={at={(0.03,0.03)},anchor=south west,legend cell align=left,align=left,draw=white!15!black,legend columns=2}
]
\addplot [color=mycolor1,solid,line width=1.0pt,mark=o,mark options={solid}]
  table[row sep=crcr]{%
2	-7.05177787096621\\
4	-7.02214460030632\\
8	-6.99851268510401\\
16	-6.99362112485536\\
32	-6.99101791002433\\
64	-6.99030273845575\\
128	-6.98981388031957\\
256	-6.98973006693999\\
512	-6.98970432145367\\
1024	-6.989702219679\\
};\addlegendentry{$\Np = 2$};
\addplot [color=mycolor2,solid,line width=1.0pt,mark size=2.5pt,mark=x,mark options={solid}]
  table[row sep=crcr]{%
4	-5.10697469260289\\
8	-4.90950136241734\\
16	-4.86918964861295\\
32	-4.78856481506665\\
64	-4.77061710961479\\
128	-4.78501618836958\\
256	-4.78501618836958\\
512	-4.78501618836958\\
1024	-4.78501618836958\\
};\addlegendentry{$\Np = 4$};
\addplot [color=mycolor3,solid,line width=1.0pt,mark size=1.5pt,mark=*,mark options={solid}]
  table[row sep=crcr]{%
8	-3.44879538876656\\
16	-3.2763060419902\\
32	-3.1454668142369\\
64	-3.05858608961668\\
128	-3.04623976166226\\
256	-3.04623976166226\\
512	-3.04623976166226\\
1024	-3.04623976166226\\
};\addlegendentry{$\Np = 8$};

\addplot [color=mycolor2,dashed,mark=triangle,mark options={solid},forget plot]
  table[row sep=crcr]{%
4	-4.77121254719662\\
8	-4.77121254719662\\
16	-4.77121254719662\\
32	-4.77121254719662\\
64	-4.77121254719662\\
128	-4.77121254719662\\
256	-4.77121254719662\\
512	-4.77121254719662\\
1024	-4.77121254719662\\
};
\addplot [color=mycolor2,dashed,mark=triangle,mark options={solid,rotate=180},forget plot]
  table[row sep=crcr]{%
4	-5.38307969816361\\
8	-5.13398661653955\\
16	-4.93106223520815\\
32	-4.82211688050969\\
64	-4.78302945612419\\
128	-4.77387384016934\\
256	-4.77198676580314\\
512	-4.77134997051686\\
1024	-4.7712658721646\\
};
\addplot [color=mycolor3,dashed,mark=triangle,mark options={solid},forget plot]
  table[row sep=crcr]{%
8	-3.01029995663981\\
16	-3.01029995663981\\
32	-3.01029995663981\\
64	-3.01029995663981\\
128	-3.01029995663981\\
256	-3.01029995663981\\
512	-3.01029995663981\\
1024	-3.01029995663981\\
};
\addplot [color=mycolor3,dashed,mark=triangle,mark options={solid,rotate=180},forget plot]
  table[row sep=crcr]{%
8	-4.46953852661122\\
16	-3.88972616756061\\
32	-3.48324074388924\\
64	-3.23927233353442\\
128	-3.10336955193267\\
256	-3.02049629965635\\
512	-3.01420523972649\\
1024	-3.01105402896064\\
};

\end{axis}
\end{tikzpicture}%
}
 \caption{Network with $\Nb = 8$. }
\label{fig: NMSE_Sc1_M=8}
\end{subfigure}
\hspace{1cm}
\begin{subfigure}[b]{0.39\columnwidth}
  {\footnotesize 
%
%
\definecolor{mycolor1}{rgb}{0.30588,0.39608,0.58039}%
\definecolor{mycolor2}{rgb}{0.85000,0.32500,0.09800}%
\definecolor{mycolor3}{rgb}{0.00000,0.74902,0.74902}%

\begin{tikzpicture}

\begin{axis}[%
width=1\textwidth,
height=1\textwidth,
at={(0\textwidth,0\textwidth)},
scale only axis,
xmode=log,
xmin=8,
xmax=1024,
xminorticks=true,
yticklabel pos=right,
xlabel={$\Nu$},
ymin=-28,
ymax=-20,
axis background/.style={fill=white}
]
\addplot [color=mycolor1,solid,line width=1.0pt,mark=o,mark options={solid}]
  table[row sep=crcr]{%
2	-27.1011736511182\\
4	-27.1011736511182\\
8	-27.1011736511182\\
16	-27.1011736511182\\
32	-27.1011736511182\\
64	-27.1011736511182\\
128	-27.1011736511182\\
256	-27.1011736511182\\
512	-27.1011736511182\\
1024	-27.1011736511182\\
};

\addplot [color=mycolor2,solid,line width=1.0pt,mark size=2.5pt,mark=x,mark options={solid}]
  table[row sep=crcr]{%
4	-24.099331233313\\
8	-24.099331233313\\
16	-24.099331233313\\
32	-24.099331233313\\
64	-24.099331233313\\
128	-24.099331233313\\
256	-24.099331233313\\
512	-24.099331233313\\
1024	-24.099331233313\\
};

\addplot[color=mycolor3,solid,line width=1.0pt,mark size=1.5pt,mark=*,mark options={solid}]
  table[row sep=crcr]{%
8	-21.1058971029925\\
16	-21.1058971029925\\
32	-21.1058971029925\\
64	-21.1058971029925\\
128	-21.1058971029925\\
256	-21.1058971029925\\
512	-21.1058971029925\\
1024	-21.1058971029925\\
};

%

\end{axis}
\end{tikzpicture}%
}
 \caption{Network with $\Nb = 1024$. }
\label{fig: NMSE_Sc1_M=1024}
\end{subfigure}
 \caption{The channel estimation performance in non-precoded and uncombined pilot transmission scenario~($\so$) as a function of the number of \acp{UE} antennas $\Nu$. $\Nb = 8$ and $\Nb = 1024$ represent cellular networks with small and large number of \ac{BS} antennas, respectively. The dashed lines represent the bounds of
 Proposition~\ref{prop: bounds-on-estimation-error-S1} for the corresponding average normalized MSE curves; however, some of the bounds are tight and can not be seen in the figure.}
\label{fig: NMSE_Sc1}
\end{figure}

To numerically illustrate Proposition~\ref{prop: bounds-on-estimation-error-S1}, we simulate a network with
various number of \ac{UE} antenna elements $\Nu$, paths $\Np$, and \ac{BS} antenna elements $\Nb$. The channel model follows~\eqref{eq:geometry-based channel model}. Our simulation parameters cover wide range of use cases, including
\begin{itemize}
  \item Access layer of traditional cellular or ad~hoc networks (small $\Nu$, small $\Nb$, large $\Np$);
  \item Access layer of sub-6~GHz massive \ac{MIMO} networks  (small $\Nu$, large $\Nb$, large $\Np$); and
  \item Access layer of \ac{mmWave} networks (small $\Nu$, large $\Nb$, small $\Np$); and
  \item Backhaul layer (large $\Nu$ and $\Nb$ values, small or large $\Np$).
\end{itemize}
We draw the AoAs and AoDs independently from uniform distributions in $[-{\pi}/{3},{\pi}/{3}]$ and $[-{\pi}/{6},{\pi}/{6}]$, respectively.
The normalized MSE is averaged over 50 realizations of AoAs and AoDs and 90,000 realizations of noise and small-scale fading.
We consider $\PilotEnergy = 0$~dB and apply the normalization $\CG_k/\NoisePower=1$ to ensure that
the average received \ac{SNR} at the BS can be described by the pilot energy and the number of antenna elements, e.g., $\text{SNR}={\PilotEnergy}/{\Nu}$ in $\so$.

Fig.~\ref{fig: NMSE_Sc1} illustrates the average normalized MSE, $e_k^{(\so)}$ as defined in Proposition~\ref{prop: bounds-on-estimation-error-S1}, against the number of antenna elements at UE $k$ for $M=8,1024$ (as small and large numbers) and three different number of paths between the UE and the BS, namely $\Np = 2,4,8$.
These numbers of multipath components cover both sparse scattering environments like in \ac{mmWave} networks and rich scattering environments like in sub-6~GHz networks.
Fig.~\ref{fig: NMSE_Sc1} shows that increasing $\Nu$ does not improve the channel estimation performance of $\so$.
The reason for this becomes clear by noting that the MMSE channel estimation error is proportional to the dimension of the received signal vector, i.e., $\vect\left(\mathbf{Y}_{\tau}^{(\so)}\right)\in \mathbb{C}^{K\Nu\Nb}$, as well as the received SNR.
On one hand, increasing $N$ linearly increases the dimension of the received signal. These extra observations, while having a constant number of unknown parameters, generally improve the channel estimation performance. On the other hand, increasing $N$ reduces the received SNR, formulated in the previous paragraph. These two effects cancel each other, making the average normalized MSE almost independent of the number of UE antennas $N$.
On the contrary, comparing Fig.~\ref{fig: NMSE_Sc1_M=8} and Fig.~\ref{fig: NMSE_Sc1_M=1024} shows that adding more antenna elements to the BS enhances the channel estimation performance. This can also be related to the increase in the dimension of the received signal by increasing the number of BS antennas, $\Nb$.
Another observation from Fig.~\ref{fig: NMSE_Sc1} is that the channel estimation performance improves for networks with sparser scattering environments (smaller $\Np$),
where this improvement is almost linear in the large antenna regimes; see Corollary~\ref{corollary: NMSE_s1}.



\subsection{Precoded and Uncombined Pilot Transmission ($\st$)}
In the second pilot transmission scenario, the pilots are precoded using spatial filters at the UEs.
However, no pilot combining is performed at the BS.
The channel estimation quality in this scenario depends on the statistical information of the channel available prior to pilot transmission.
Assuming that $\Au_k$ is available at UE $k$, either perfectly or with some unbiased errors\cite{36.214},
the spatial filters can be designed to help focusing the pilot energy
along the strongest multipath components
between the UEs and the BS.
Hence, precoding the pilots boosts the SNR in the pilot transmission phase, which is specially beneficial for the UEs at the cell edges.

Considering that there are $\Np \leq \Nu$ paths between each UE and the BS, $T_\tau^{(\st)} = K \Np \le T_\tau^{(\so)}$ pilot symbols
suffices for transmitting orthogonal training sequences through all the paths.\footnote{It is shown in~\cite{Bjornson2010}
that in the case of correlated channels between UE $k$ and the BS,
the actual length of training sequence needed for channel estimation can be less than the number of transmitting antennas.}
In other words, unlike scenario $\so$, where orthogonal pilots are assigned to the UE antennas,
the number of required orthogonal pilot sequences in scenario $\st$ is equal to the number of multipath components.
Clearly, this pilot transmission scheme leaves longer time for the data transmission phase, compared to the baseline scenario $\so$.
This brings a significant gain for the data transmission time, especially in wireless networks with small coherence time such as mmWave networks~\cite{shokri2015mmWavecellular}.

Let $\mathbf{P}_{k}^{(\st)}\in \mathbb{C}^{\Np\times T_\tau^{(\st)} }$ be the pilot symbols transmitted by UE $k$ in scenario $\st$, then orthogonality of training sequences and the energy constraint \eqref{eq: pilot energy constraint} imply that
\begin{equation}\label{eq: pilot matrix_s2}
\mathbf{P}_{k}^{(\st)}\left(\mathbf{P}_{j}^{(\st)}\right)\herm =
\begin{cases}
 \frac{\PilotEnergy}{\Np} \mathbf{I}_{\Np} & k = j \:, \\
 \boldsymbol{0} & \mbox{otherwise} \:.
\end{cases}
\end{equation}

Note that although in this paper we are investigating the uplink channel estimation,
pilot transmission scenario $\st$ entails the same complexity for downlink channel estimation.
In contrast, the complexity of the downlink channel estimation in scenario $\so$ is substantially
higher than that of uplink channel estimation if $\Nb \gg K \Nu$.
As a result, in massive MIMO systems,
the $\so$ scheme
is suitable only for the \ac{TDD} mode
(in which the channel reciprocity principle holds),
whereas $\st$ can be used in both \ac{TDD} and \ac{FDD} modes.
Note that the number of unique pilots required by $\st$ does still scale by the number of \acp{UE}.
Moreover, due to the lack of receiver antenna gain,
it may not completely solve the imbalanced pilot-data coverage problem, though substantially alleviate it compared to the $\so$ scheme.

For mathematical tractability, we assume the availability of 
of perfect second-order statistics information which allows us to
gain insights about the impact of different parameters on the network performance
(e.g., channel estimation quality and network throughput).
In this section, we choose the precoding matrix $\mathbf{V}_k = \Au_k$ for each UE $k$,
which simplifies the mathematical analysis and, at the same time, is asymptotically optimal in terms of maximizing the SNR~\cite{Ayach2012Capacity}.
No combining filter is considered at the BS in this scenario, that is $\mathbf{W}_k = \I_{\Nb}$.
Substituting the training matrix and spatial filters of scenario $\st$ into \eqref{eq: MMSE channel estimate}, the \ac{MMSE} estimate of the channel becomes
\begin{align}
\label{eq: MMSE channel estimate-S2}
\vect\left(\widehat{\mathbf{H}}_k^{(\st)}\right)
=\,&\mathbf{R}_{k} \Pt^{(\st)}_k
\left(\left(\Pt^{(\st)}_k\right)^{\mbox{\scriptsize H}}
\mathbf{R}_{k} \Pt^{(\st)}_k + \NoisePower \I\right)^{-1}\nonumber\\
&\cdot \vect\left(\mathbf{Y}_{\tau}^{(\st)}\right),
\end{align}
where $\Pt^{(\st)}_k=\big(\Au_k\mathbf{P}_k^{(\st)}\big)^* \otimes \I_{\Nb}$.
$\{\Pt^{(\st)}_k\}_{k=1}^{K}$ inherit
the orthogonality property of $\{\mathbf{P}^{(\st)}_k\}_{k=1}^{K}$, similar to $\so$,
and therefore the received signals from UE $j\ne k$ are canceled out at the receiver in the estimation process of $\mathbf{H}_k$.

Define the estimation error matrix as $\Ht_k^{(\st)} = \mathbf{H}_k - \Hh_k^{(\st)}$.
The following propositions characterize the accuracy of the channel estimation under precoded but uncombined pilot transmission:
\begin{prop} \label{prop: covariance of error_s2}
Consider the system model of scenario~$\st$.
The covariance of the channel estimation error matrix using orthogonal pilot transmissions
given by~\eqref{eq: pilot matrix_s2}
and an MMSE estimator given by~\eqref{eq: MMSE channel estimate-S2} is
\be
\label{eq: covariance of error_s2} 
\begin{split}
{\Rt}_{k}^{(\st)} \!\!&=
\E\left[\vect\left(\widetilde{\mathbf{H}}^{(\st)}_k\right) \vect\left(\widetilde{\mathbf{H}}^{(\st)}_k\right)\herm\right] \\
&=\delta \Nb \CG_k \left(\Au_k^* \odot \Ab_k \right)  \\
&~~~\cdot \left(\I_{\Np} \!+\!  \delta\Nb\zeta_k \left(\mathbf{R}_{\Au_k}^2\right)\tran\!\!\circ\!\mathbf{R}_{\Ab_k} \!\right)^{\!\!-1}
\!\!\left(\Au_k^*\!\odot\!\Ab_k \right)\herm.
\end{split}
\ee
\end{prop}
\begin{IEEEproof}
A proof is given in Appendix~B.
\end{IEEEproof}

\begin{prop}\label{prop: bounds-on-estimation-error-S2}
Consider the covariance of the channel estimation error for UE $k$ given by Proposition~\ref{prop: covariance of error_s2}. The normalized \ac{MSE}, defined as $e_k^{(\st)} = \tr(\Rt_k^{(\st)})/\tr(\mathbf{R}_k)$, is bounded as
\be\label{eq: NMSE_s2}
\frac{\lambda_{\max}^{-1}}{1\!+\!\delta\Nb\zeta_k}\! \left[1 \!-\! \frac{\epsilon_k^{(\st)}}
{\delta\Nb\zeta_k}\right]^{\!+}
\!\!\!\le\! e_k^{(\st)} \!\le\!
\frac{\lambda_{\min}^{-1}}{1\!+\!\delta\Nb\zeta_k},
\ee
where $\lambda_{\min}$ and $\lambda_{\max}$ represent the minimum and maximum eigenvalues of $\mathbf{R}_{\Au_k}$, respectively and
\be
\epsilon_k^{(\st)} = \frac{\left(1-\cond\left(\I_{\Np} +  \delta\Nb\zeta_k \left(\mathbf{R}_{\Au_k}^2\right)\tran \circ \mathbf{R}_{\Ab_k} \right)\right)^2}
{4\cond\left(\I_{\Np} +  \delta\Nb\zeta_k \left(\mathbf{R}_{\Au_k}^2\right)\tran \circ \mathbf{R}_{\Ab_k} \right)} \ge 0.
\ee
\end{prop}

\begin{corollary}\label{corollary: NMSE_s2}
As $\Nu \to \infty$ (so $\mathbf{R}_{\Au_k} \to \I_{\Np}$), both upper and lower bounds in~\eqref{eq: NMSE_s2} become tighter, and ${e_k^{(\st)} \to (1+\delta \Nb\zeta_k)^{-1}}$.
\end{corollary}
Corollaries~\ref{corollary: NMSE_s1} and~\ref{corollary: NMSE_s2} characterize the estimation error in scenarios~$\so$ and~$\st$, respectively. In particular, when $\Nu \to \infty$, pilot precoding can improve the channel estimation error by a factor of
\be
\label{eq: channel estimation gain}
\lim_{N \to \infty} \frac{e_k^{(\st)}}{e_k^{(\so)}}
= \dfrac{1+\Nb\zeta_k}{1+ \delta\Nb\zeta_k}
\geq \frac{1}{\delta} \:,
\ee
where the equality holds if $\Nb \zeta_k$ is sufficiently large.
The proof of~\eqref{eq: channel estimation gain} is a straightforward application of Lemma~\ref{lemma: inequality_for_delta} in Appendix~A.

\begin{figure}
\begin{subfigure}[b]{\columnwidth}
 \centering
  {\footnotesize 
%
%
\definecolor{mycolor1}{rgb}{0.30588,0.39608,0.58039}%
\definecolor{mycolor2}{rgb}{0.85000,0.32500,0.09800}%
\definecolor{mycolor3}{rgb}{0.00000,0.74902,0.74902}%
\begin{tikzpicture}

\begin{axis}[%
width=0.85\columnwidth,
height=0.425\columnwidth,
at={(0\textwidth,0\textwidth)},
scale only axis,
xmode=log,
xmin=8,
xmax=1024,
xminorticks=true,
xlabel={Number of UE antenna elements $N$},
ymin=-35,
ymax=0,
ytick = {-35,-30,-25,-20,-15,-10,-5,0},
ylabel={Average normalized MSE (dB)},
axis background/.style={fill=white},
legend style={at={(0.03,0.03)},anchor=south west,legend cell align=left,align=left,draw=white!15!black}
]

\addplot [color=mycolor1,solid,line width=1.0pt,mark=o,mark options={solid}]
  table[row sep=crcr]{%
2	-9.18959339406993\\
4	-11.226615160274\\
8	-13.3817498239453\\
16	-15.533423883738\\
32	-18.2311174235457\\
64	-21.1922050999713\\
128	-24.0636422144247\\
256	-27.058522855527\\
512	-30.1004009986772\\
1024	-33.0201142015496\\
};
\addlegendentry{$\Np = 2$};

\addplot [color=mycolor2,solid,line width=1.0pt,mark size=2.5pt,mark=x,mark options={solid}]
  table[row sep=crcr]{%
4	-8.14508828574081\\
8	-9.50501268558472\\
16	-10.9345241987045\\
32	-12.9828943073883\\
64	-15.4721046642081\\
128	-18.2091795258359\\
256	-21.0969781786436\\
512	-24.0740349517937\\
1024	-27.0712588765868\\
};
\addlegendentry{$\Np=4$};

\addplot [color=mycolor3,solid,line width=1.0pt,mark size=1.5pt,mark=*,mark options={solid}]
  table[row sep=crcr]{%
8	-6.63466196030581\\
16	-7.30469731926928\\
32	-8.54467639007594\\
64	-10.3320011519483\\
128	-12.5717410584418\\
256	-15.20500890368\\
512	-18.1455328364369\\
1024	-21.0958912017351\\
};
\addlegendentry{$\Np=8$};

\addplot [color=mycolor1,dashed,mark=triangle,mark options={solid}]
  table[row sep=crcr]{%
2	-0.00625236479579538\\
4	-1.80841230101084\\
8	-5.73722990268663\\
16	-10.3100450250974\\
32	-14.5213902938936\\
64	-19.898860476267\\
128	-23.7983825440811\\
256	-27.0322810119203\\
512	-30.0721933934337\\
1024	-33.0953956074537\\
};

\addplot [color=mycolor1,dashed,mark=triangle,mark options={solid,rotate=180}]
  table[row sep=crcr]{%
2	-9.98721300375232\\
4	-11.9473860141624\\
8	-13.8809044284056\\
16	-16.0891989062713\\
32	-18.6632626452871\\
64	-21.4169167415378\\
128	-24.2584150032041\\
256	-27.1627046811576\\
512	-30.1411088029426\\
1024	-33.1348071238527\\
};

\addplot [color=mycolor2,dashed,mark=triangle,mark options={solid}]
  table[row sep=crcr]{%
4	0\\
8	0\\
16	-0.662992711358033\\
32	-4.30743378282842\\
64	-9.42279523943398\\
128	-16.3274442924381\\
256	-20.6670764099528\\
512	-23.8956465514782\\
1024	-26.996685982998\\
};

\addplot [color=mycolor2,dashed,mark=triangle,mark options={solid,rotate=180}]
  table[row sep=crcr]{%
4	-12.2811396134731\\
8	-12.171389060952\\
16	-12.858919731421\\
32	-14.3140180023776\\
64	-16.3405934415535\\
128	-18.7248600384445\\
256	-21.4169547214426\\
512	-24.2755762557982\\
1024	-27.1975390606704\\
};

\addplot [color=mycolor3,dashed,mark=triangle,mark options={solid}]
  table[row sep=crcr]{%
8	0\\
16	0\\
32	-0.0181352655322658\\
64	-1.25315987214613\\
128	-7.10820873421634\\
256	-14.1176585552342\\
512	-17.5881945523296\\
1024	-20.8526458164214\\
};

\addplot [color=mycolor3,dashed,mark=triangle,mark options={solid,rotate=180}]
  table[row sep=crcr]{%
8	-18.4992349267152\\
16	-13.7204870189833\\
32	-12.008342900386\\
64	-12.5560768839742\\
128	-14.0622845574947\\
256	-15.9771110798918\\
512	-18.5868026144781\\
1024	-21.3249892402326\\
};

%
%
%
%

\end{axis}
\end{tikzpicture}%
}
 \caption{Network with $\Nb = 8$.  }
\label{fig: NMSE_Sc2_M=8}
\end{subfigure}
\begin{subfigure}[b]{\columnwidth}
 \centering
  {\footnotesize 
%
%
\definecolor{mycolor1}{rgb}{0.30588,0.39608,0.58039}%
\definecolor{mycolor2}{rgb}{0.85000,0.32500,0.09800}%
\definecolor{mycolor3}{rgb}{0.00000,0.74902,0.74902}%
\begin{tikzpicture}

\begin{axis}[%
width=0.85\columnwidth,
height=0.425\columnwidth,
at={(0\textwidth,0\textwidth)},
scale only axis,
unbounded coords=jump,
xmode=log,
xmin=8,
xmax=1024,
xminorticks=true,
xlabel={Number of UE antenna elements $\Nu$},
ymin=-55,
ymax=-20,
ytick = {-55,-50,-45,-40,-35,-30,-25,-20},
ylabel={Average normalized MSE (dB)},
axis background/.style={fill=white},
legend style={at={(0.03,0.03)},anchor=south west,legend cell align=left,align=left,draw=white!15!black}
]
\addplot  [color=mycolor1,solid,line width=1.0pt,mark=o,mark options={solid}]
  table[row sep=crcr]{%
2	-29.7886139968648\\
4	-32.0034174819494\\
8	-34.2391000098963\\
16	-36.5909218911391\\
32	-39.1947525612139\\
64	-42.3013099081847\\
128	-44.8536831945009\\
256	-48.0959729346125\\
512	-51.1400871363058\\
1024	-54.1653918099465\\
};\addlegendentry{$\Np = 2$};
\addplot [color=mycolor2,solid,line width=1.0pt,mark size=2.5pt,mark=x,mark options={solid}]
  table[row sep=crcr]{%
4		-28.465954961037720\\
8		-30.010708182421311\\
16		-31.750666021273215\\
32		-33.899856936497031\\
64		-36.2737700970469\\
128	-38.831018728303948\\
256	-42.080530937924252\\
512	-45.140907444597197\\
1024		-48.160803524299602\\
};\addlegendentry{$\Np = 4$};
\addplot [color=mycolor3,solid,line width=1.0pt,mark size=1.5pt,mark=*,mark options={solid}]
  table[row sep=crcr]{%
8		-26.243601070867378\\
16		-27.338837812820689\\
32		-28.968737478893086\\
64		-31.090883050634236\\
128	-33.490007611271864\\
256	-36.084555699761700\\
512	-39.092344530155117\\
1024	 -42.363465578206345\\
};\addlegendentry{$\Np = 8$};

\addplot[color=mycolor1,dashed,mark=triangle,mark options={solid}]
  table[row sep=crcr]{%
2	-7.30716161636336\\
4	-12.3460972490271\\
8	-19.4469076683064\\
16	-28.1232593255695\\
32	-35.526686276059\\
64	-40.9374278304584\\
128	-44.8536831945009\\
256	-48.0959729346125\\
512	-51.1400871363058\\
1024	-54.1653918099465\\
};
\addplot  [color=mycolor1,dashed,mark=triangle,mark options={solid,rotate=180}]
  table[row sep=crcr]{%
2	-29.9824517767971\\
4	-32.4560243488856\\
8	-34.6855730576852\\
16	-37.0272778913611\\
32	-39.6683573318262\\
64	-42.4554389607674\\
128	-45.3137075810315\\
256	-48.2263954013952\\
512	-51.2090024449869\\
1024	-54.2048033007314\\
};

\addplot[color=mycolor2,dashed,mark=triangle,mark options={solid}]
  table[row sep=crcr]{%
4	0\\
8	-1.0386217107249\\
16	-6.6923507926258\\
32	-13.4231241721107\\
64	-28.2607319883464\\
128	-37.3327402746035\\
256	-41.7056437641443\\
512	-44.9509472018979\\
1024	-48.0603779056902\\
};
\addplot [color=mycolor2,dashed,mark=triangle,mark options={solid,rotate=180}]
  table[row sep=crcr]{%
4	-29.2864314085284\\
8	-31.3004727745604\\
16	-33.1202544008597\\
32	-35.076305162011\\
64	-37.2737700970469\\
128	-39.7292971820044\\
256	-42.4554181117042\\
512	-45.3308676872965\\
1024	-48.261229142909\\
};
\addplot [color=mycolor3,dashed,mark=triangle,mark options={solid}]
  table[row sep=crcr]{%
8	0\\
16	0\\
32	-2.10922785776922\\
64	-11.6318388725423\\
128	-27.812318055212\\
256	-35.0571787959867\\
512	-38.593490534495\\
1024	-41.8912131706128\\
};
\addplot [color=mycolor3,dashed,mark=triangle,mark options={solid,rotate=180}]
  table[row sep=crcr]{%
8	-27.6094906469546\\
16	-29.1996618023466\\
32	-30.8998660110254\\
64	-32.7590018330438\\
128	-34.7806052305278\\
256	-36.9119326035367\\
512	-39.5911985258152\\
1024	-42.3634655782063\\
};
\end{axis}
\end{tikzpicture}%
}
 \caption{Network with $\Nb = 1024$. }
\label{fig: NMSE_Sc2_M=1024}
\end{subfigure}
 \caption{The channel estimation performance in the precoded uncombined pilot transmission scenario~($\st$) as a function of the number of \acp{UE} antenna elements. $\Nb = 8$ and $\Nb = 1024$ represent cellular networks with small and large number of \ac{BS} antennas, respectively.
 The dashed lines represent the bounds of Proposition~\ref{prop: bounds-on-estimation-error-S2} for the corresponding average normalized MSE curves.}
\label{fig: NMSE_Sc2}
\end{figure}

To numerically evaluate the performance of channel estimation using $\st$,
we use the same simulation setting as the one used in Fig.~\ref{fig: NMSE_Sc1}.
The average normalized MSE and the corresponding bounds,
as computed in \eqref{eq: NMSE_s2}, against the number of antenna elements at the \acp{UE} are illustrated in Fig.~\ref{fig: NMSE_Sc2} for three different values of $\Np$, namely $\Np = 2,4,8$.
As the figure shows, unlike $\so$, increasing $\Nu$ significantly boosts the channel estimation performance.
Moving from $\Nu = 8$ to $\Nu = 1024$ reduces the estimation error by around 20~dB.
The reason is that
in $\st$, as $\Nu$ grows large, the energy transmitted through the paths between the UEs and the BS increases.
Therefore, unlike $\so$, the received SNR at the BS in $\st$ increases with $\Nu$.
Again, similarly to $\so$, by comparing Fig.~\ref{fig: NMSE_Sc2_M=8} and Fig.~\ref{fig: NMSE_Sc2_M=1024},
we can see that when $\Nb$ increases from 8 to 1024, a gain of $10\log_{10}({1024}/{8}) \approx 21$~dB can be achieved.

The gain of pilot precoding in $\st$ can be understood by comparing Fig.~\ref{fig: NMSE_Sc1} and Fig.~\ref{fig: NMSE_Sc2}.
Specifically, when the number of antenna elements at both the UE and BS sides is large, say $\Nb=\Nu = 1024$,
Fig.~\ref{fig: NMSE_Sc2_M=1024} shows 21, 24 and 27~dB lower MSE compared with
Fig.~\ref{fig: NMSE_Sc1_M=1024} for the curves corresponding to $\Np = 2,4$ and $8$, respectively.
This fact is in accordance with~\eqref{eq: channel estimation gain}.

\subsection{Precoded and Combined Pilot Transmission ($\sth$)}\label{sec: PrecodedCombinedPilots}
In the third scenario, in addition to precoding the pilots at UEs, the received signals at the BS are also combined,
using the available information about the AoAs at the BS.
In this case, in addition to the gains discussed earlier for $\st$,
exploiting the spatial filters at the BS, given the large number of BS antennas, can lead to a sufficiently good spatial separation of the UEs. 
Therefore, a combiner at the BS may enable us to use non-orthogonal pilots for different UEs, if their orthogonality can be maintained in the spatial domain.
Therefore, in this scenario,
non-orthogonal sequences with $T_\tau^{(\sth)}<K\Np$ symbols are transmitted from each antenna elements.
Without loss of generality, we consider the
extreme scenario of $T_{\tau}^{\sth} = 1$ in our mathematical analysis,
that is only one pilot will be used to estimate the entire uplink channel matrix.
Due to the pilot reuse, there is a contamination of the pilots at the BS side.
This situation is similar to a multi-cell network where pilot reuse in neighboring cells causes the pilot contamination problem.
Notice that scenario $\sth$ addresses the pilot contamination problem, though
we have a single cell network setting.
We then show with numerical analysis that using $T_{\tau}^{\sth} > 1$ orthogonal pilots brings improvement in the channel estimation performance
at the expense of having less time for the data transmission phase.
Notice that, similarly to $\st$, pilot transmission scenario $\sth$ enables
the realization of massive MIMO using both \ac{TDD} and \ac{FDD} schemes,
as the channel estimation complexity grows only with the number of paths,
rather than with the number of antennas, which is a useful property for both UL and DL pilots.
Altogether, $\sth$ with $1 \leq T_{\tau}^{\sth} <K\Np$ pilots
is a promising option that not only scales well with the number of \acp{UE}
but also eliminates the imbalanced pilot-data coverage problem.

We assume that $\Au_k$  is available at UE $k$ and all the $\Au_k$'s and $\Ab_k$'s are available at the BS either perfectly or with some unbiased error.
In the rest of this section for the sake of mathematical tractability,
we assume that the knowledge about the AoAs and AoDs at the BS and UEs is perfect.
Later, using numerical simulations,
we investigate the effect of imperfect knowledge about AoAs and AoDs on the performance of the network.
Exploiting the available information about the channel, similarly
to $\st$, the pilot precoding filter at UE $k$ is designed as $\mathbf{V}_k = \Au_k$.
Moreover,
recalling that the main objective of this paper is to characterize the gains
of pilot precoding/combining rather than finding the optimal precoders and combiners,
we design the combining filter for UE $k$ as $\mathbf{W}_k = \Ab_k$.

Substituting for the spatial filters in \eqref{eq: vectorized received pilot}, the received signal in scenario $\sth$ is
\be
\label{eq: received signal-s3}
\begin{split}
\vect\!\left(\mathbf{Y}^{(\sth)}_{\tau,k}\right) \!&=\! \left(\!\Pt^{(\sth)}_{k}\!\right)\herm\!\!\vect\left(\mathbf{H}_k\right)\!+\!
\left(\I \!\otimes\! \Ab_k\herm \right)\vect\left(\mathbf{Z}_\tau \right) \\
&~~~+\underbrace{\sum\limits_{j=1,j\neq k}^{K}\! \left(\!\Pt^{(\sth)}_{kj}\!\right)\herm\!\!\vect\left(\mathbf{H}_j\right)}_{\text{Inter-UE Interference}}\:,
\end{split}
\ee
where $\big(\Pt^{(\sth)}_{kj}\big)\herm = \big(\Au_j \mathbf{P}_j^{(\sth)}\big)\tran \otimes \Ab_k\herm$
with $\mathbf{P}^{(\sth)}_j \in \mathbb{C}^{\Np\times T_\tau^{(\sth)}}$ being the pilot matrix transmitted from UE $j$.

Unlike the two previous scenarios, in $\sth$, \emph{pilot contamination} is inevitable due to non-orthogonal pilot transmissions.
In fact, pilot contamination may contain two parts: the interference from pilots transmitted from the antenna elements of the same UE, called \emph{intra-UE interference}, and the interference from the pilots transmitted from other UEs, called \emph{inter-UE interference}.
Note that, in the case of multicell networks, the interference from the UEs in other cells, called \emph{inter-cell interference}, can also contaminate the received pilots.
The pilot precoding used in this scenario is beneficial for boosting the link budget and reducing
the inter-UE interference while the pilot combining mitigates both inter-UE and inter-cell interferences.


Define the covariance matrix of the inter-UE interference term and the covariance of the received signal without inter-UE interference respectively by
\begin{align}
\label{eq: Q-bar}
\widebar{\mathbf{Q}}_k &= \sum\limits_{j=1,j\neq k}^{K} \left(\Pt^{(\sth)}_{kj}\right)^{\!\!\mbox{\scriptsize H}}
\mathbf{R}_{j} \Pt^{(\sth)}_{kj} \; ,\\
\mathbf{Q}_k &=  \left(\Pt^{(\sth)}_{k}\right)^{\!\!\mbox{\scriptsize H}}
\mathbf{R}_{k} \Pt^{(\sth)}_{k} + \NoisePower  \left(\I \otimes \mathbf{R}_{\Ab_k} \right).
\end{align}
It is now straightforward
to show that the MMSE estimate of the vectorized channel in this scenario is
\be
\label{eq: MMSE-estimate-S3}
\begin{split}
\vect\left(\widehat{\mathbf{H}}_k^{(\sth)}\right)
= \mathbf{R}_{k} \Pt^{(\sth)}_{k}
\left(\mathbf{Q}_k + \widebar{\mathbf{Q}}_k\right)^{-1}\vect\left(\mathbf{Y}^{(\sth)}_{\tau,k}\right).
\end{split}
\ee
Let $\Ht_k^{(\sth)} = \mathbf{H}_k - \Hh_k^{(\sth)}$ be the error due to estimating channel $\vec{H}_k$ in the $\sth$ scenario.
Then,
\be
\begin{split}
\label{eq:EstErrCov}
\Rt_{k}^{(\sth)} &= \E\left[\vect\left(\Ht_k^{(\sth)}\right)\vect\left(\Ht_k^{(\sth)}\right)\herm\right]\\
&=\mathbf{R}_{k} -
\mathbf{R}_{k}\Pt^{(\sth)}_{k}\left(\mathbf{Q}_k + \widebar{\mathbf{Q}}_k\right)^{-1}\left(\Pt^{(\sth)}_{k}\right)\herm \mathbf{R}_{k} \:.
\end{split}
\ee

\begin{figure}
\begin{subfigure}[b]{\columnwidth}
\centering
{\footnotesize 
%
%
\definecolor{mycolor1}{rgb}{0.30588,0.39608,0.58039}%
\definecolor{mycolor2}{rgb}{0.85000,0.32500,0.09800}%
\definecolor{mycolor3}{rgb}{0.00000,0.74902,0.74902}%
\begin{tikzpicture}

\begin{axis}[%
width=0.85\columnwidth,
height=0.425\columnwidth,
at={(0\textwidth,0\textwidth)},
scale only axis,
xmode=log,
xmin=8,
xmax=1024,
xminorticks=true,
xlabel={Number of UE antenna elements N},
ymin=-35,
ymax=0,
ylabel={Average normalized MSE (dB)},
axis background/.style={fill=white},
legend style={at={(0.01,0.01)},anchor=south west,legend cell align=left,align=left,draw=white!15!black,legend columns=2},
ytick={-35,-30,-25,-20,-15,-10,-5,0},
]

\addplot [color=mycolor1,solid,line width=1.0pt,mark options={solid}]table[row sep=crcr]{%
2	-8.4428945585914\\};\addlegendentry{$\Np=2$};
\addplot [color=white,solid,line width=1.0pt,mark=o,mark options={mycolor1,solid}]table[row sep=crcr]{%
4	-8.14508828574081\\};\addlegendentry{$\Tt=1$};
\addplot [color=mycolor2,dashed,line width=1.0pt,mark options={solid}]table[row sep=crcr]{%
4	-5.55648967450095\\};\addlegendentry{$\Np=4$};
\addplot [color=white,solid,line width=1.0pt,mark size = 1.5pt,mark=*,mark options={mycolor1,solid}]table[row sep=crcr]{%
2	-9.3180646323645\\};\addlegendentry{$\Tt=\Np$};
\addplot [color=mycolor3,dashdotted,line width=1.0pt,mark options={solid}]table[row sep=crcr]{%
8	-3.17819544689713\\};\addlegendentry{$\Np=8$};

\addplot [color=mycolor1,solid,line width=1.0pt,mark=o,mark options={solid}]
  table[row sep=crcr]{%
2	-8.4428945585914\\
4	-8.14244939522043\\
8	-8.01975681169944\\
16	-8.2846164017668\\
32	-8.89722539099743\\
64	-9.14707561527284\\
128	-9.09180194289448\\
256	-9.36349258790823\\
512	-9.35211093686076\\
1024	-9.37872413347156\\
};

\addplot [color=mycolor2,dashed,line width=1.0pt,mark=o,mark options={solid}]
  table[row sep=crcr]{%
4	-5.55648967450095\\
8	-5.18316653311036\\
16	-4.67925910038664\\
32	-4.67635807833185\\
64	-4.89083057908112\\
128	-5.02799101472863\\
256	-5.10857858072949\\
512	-5.13432601288011\\
1024	-5.17215319139759\\
};

\addplot [color=mycolor3,dashdotted,line width=1.0pt,mark=o,mark options={solid}]
  table[row sep=crcr]{%
8	-3.17819544689713\\
16	-2.68022271133217\\
32	-2.60821171259406\\
64	-2.60043526086778\\
128	-2.67131328224594\\
256	-2.72047189225988\\
512	-2.85738284519196\\
1024	-2.93794410666297\\
};

\addplot [color=mycolor1,solid,line width=1.0pt,mark size = 1.5pt,mark=*,mark options={solid}]
  table[row sep=crcr]{%
2	-9.3180646323645\\
4	-11.226615160274\\
8	-13.3817498239453\\
16	-15.533423883738\\
32	-18.2311174235457\\
64	-21.1922050999713\\
128	-24.0636422144247\\
256	-27.0585228555271\\
512	-30.1004009986772\\
1024	-33.0201142015496\\
};

\addplot [color=mycolor2,dashed,line width=1.0pt,mark size = 1.5pt,mark=*,mark options={solid}]
  table[row sep=crcr]{%
4	-8.14508828574081\\
8	-9.50501268558472\\
16	-10.9345241987045\\
32	-12.9828943073884\\
64	-15.4721046642081\\
128	-18.2091795258359\\
256	-21.0969781786436\\
512	-24.0740349517937\\
1024	-27.0712588765869\\
};

\addplot [color=mycolor3,dashdotted,line width=1.0pt,mark size = 1.5pt,mark=*,mark options={solid}]
  table[row sep=crcr]{%
8	-6.63466196030573\\
16	-7.30469731926917\\
32	-8.5446763900762\\
64	-10.3320011519478\\
128	-12.5717410584388\\
256	-15.2050089036798\\
512	-18.1455328364341\\
1024	-21.0958912017386\\
};

\end{axis}
\end{tikzpicture}%
}
\caption{Network with $\Nb = 8$.}
\label{fig: NMSE_Sc3_M=8}
\end{subfigure}
\begin{subfigure}[b]{\columnwidth}
  {\footnotesize 
%
%
\definecolor{mycolor1}{rgb}{0.30588,0.39608,0.58039}%
\definecolor{mycolor2}{rgb}{0.85000,0.32500,0.09800}%
\definecolor{mycolor3}{rgb}{0.00000,0.74902,0.74902}%
\begin{tikzpicture}

\begin{axis}[%
width=0.85\columnwidth,
height=0.425\columnwidth,
at={(0\textwidth,0\textwidth)},
scale only axis,
xmode=log,
xmin=8,
xmax=256,
xminorticks=true,
xlabel={Number of UE antenna elements N},
ymin=-50,
ymax=-25,
ylabel={Average normalized MSE (dB)},
axis background/.style={fill=white},
legend style={at={(0.01,0.01)},anchor=south west,legend cell align=left,align=left,draw=white!15!black,legend columns=2},
ytick={-50,-45,-40,-35,-30,-25},
]
\addplot [color=mycolor1,solid,line width=1.0pt,mark options={solid}]table[row sep=crcr]{%
2	-8.4428945585914\\};\addlegendentry{$\Np=2$};
\addplot [color=white,solid,line width=1.0pt,mark=o,mark options={mycolor1,solid}]table[row sep=crcr]{%
4	-8.14508828574081\\};\addlegendentry{$\Tt=1$};
\addplot [color=mycolor2,dashed,line width=1.0pt,mark options={solid}]table[row sep=crcr]{%
4	-5.55648967450095\\};\addlegendentry{$\Np=4$};
\addplot [color=white,solid,line width=1.0pt,mark size = 1.5pt,mark=*,mark options={mycolor1,solid}]table[row sep=crcr]{%
2	-9.3180646323645\\};\addlegendentry{$\Tt=\Np$};
\addplot [color=mycolor3,dashdotted,line width=1.0pt,mark options={solid}]table[row sep=crcr]{%
8	-3.17819544689713\\};\addlegendentry{$\Np=8$};

\addplot [color=mycolor1,solid,line width=1.0pt,mark size = 1.5pt,mark=*,mark options={solid}]
  table[row sep=crcr]{%
2	-29.0253242785333\\
4	-31.7324151114674\\
8	-33.7671013862326\\
16	-35.971202850756\\
32	-37.5059237899336\\
64	-39.1947969403116\\
128	-40.4317368544306\\
256	-41.3109303952054\\
};

\addplot [color=mycolor1,solid,line width=1.0pt,mark=o,mark options={solid}]
  table[row sep=crcr]{%
2	-32.7411518659374\\
4	-33.3459474807523\\
8	-34.7564078648987\\
16	-36.3994052905504\\
32	-39.332246877274\\
64	-42.2966340650495\\
128	-45.2090078277132\\
256	-48.1653929875895\\
};

\addplot [color=mycolor2,dashed,line width=1.0pt,mark size = 1.5pt,mark=*,mark options={solid}]
  table[row sep=crcr]{%
4	-28.1969515278413\\
8	-29.8907261116925\\
16	-31.57485904173\\
32	-33.65567955316\\
64	-36.0574470164977\\
128	-38.4428325483675\\
256	-40.5001558531019\\
};

\addplot [color=mycolor2,dashed,line width=1.0pt,mark=o,mark options={solid}]
  table[row sep=crcr]{%
4	-32.1579650163835\\
8	-32.0564841977573\\
16	-32.327639596417\\
32	-34.1336616406775\\
64	-36.5631135265133\\
128	-39.3019117726271\\
256	-42.1180008286321\\
};

\addplot [color=mycolor3,dashdotted,line width=1.0pt,mark=o,mark options={solid}]
  table[row sep=crcr]{%
8	-29.0797151665536\\
16	-28.6994079999736\\
32	-29.6759112988534\\
64	-31.248167111406\\
128	-33.4459194063795\\
256	-36.1280905989571\\
};

\addplot [color=mycolor3,dashdotted,line width=1.0pt,mark size = 1.5pt,mark=*,mark options={solid}]
  table[row sep=crcr]{%
8	-26.243601070867392\\
16	-27.338837812820703\\
32	-28.968737478893086\\
64	-31.090883050634236\\
128	-33.490007611271864\\
256	-36.204797752740120\\
};

\end{axis}
\end{tikzpicture}%
}
\caption{Network with $\Nb = 1024$.}
\label{fig: NMSE_Sc3_M=1024}
\end{subfigure}
 \caption{Channel estimation performance in
 the precoded and combined pilot transmission scenario ($\sth$) as a function of the number of UEs antenna elements
 with $\Tt=\Np$ and $\Tt=1$ training lengths.
 $\Nb = 8$ and $\Nb = 1024$ represent cellular networks with small and large number of \ac{BS} antennas, respectively.}
\label{fig: NMSE_Sc3}
\end{figure}
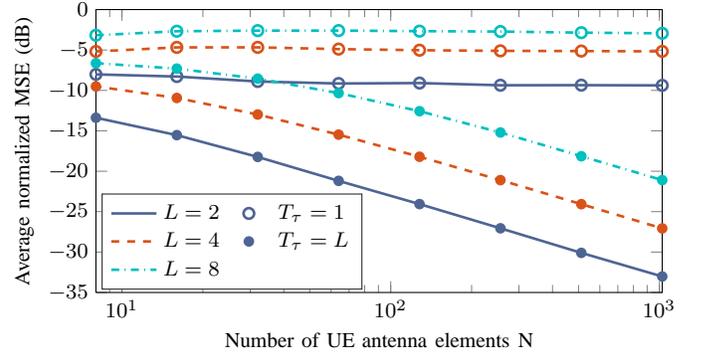
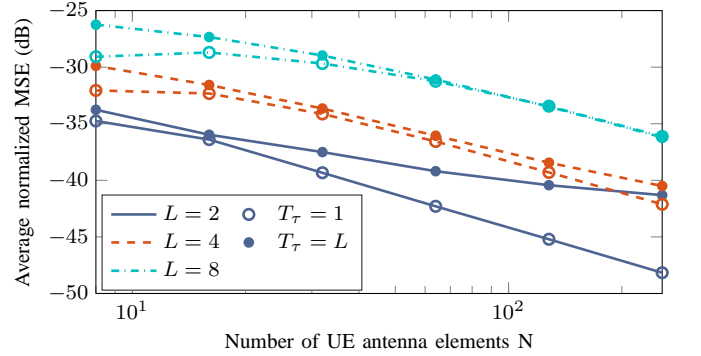

From \eqref{eq:EstErrCov}, we have the following useful corollary:
\begin{corollary}\label{corollary: pilot_decontamination}
In $\sth$, the pilot contamination caused by inter-UE and intra-UE interference tend to zero, when the number antenna elements at the BS and UEs grow large, respectively.
\end{corollary}
\begin{IEEEproof}
  A proof is given in Appendix~B.
\end{IEEEproof}
Corollary~\ref{corollary: pilot_decontamination} implies that the combiner substantially reduces the
pilot contamination term, and asymptotically makes it zero.
It is straightforward to show that Corollary~\ref{corollary: pilot_decontamination} holds also for multicell networks, and only one pilot in the asymptotic regime is enough for channel estimation in the entire network.

In the non-asymptotic case,
using more orthogonal pilots per \ac{UE} allows to reduce the number of antennas contributing to the pilot contamination term
and to maintain the \ac{UE}'s transmit power under a desirable threshold.
Finding the minimum number of orthogonal pilots for a given maximum power
of the pilot contamination is an interesting topic for future works.


We use the same simulation setting as that of Fig.~\ref{fig: NMSE_Sc1} to analyze the performance of $\sth$.
However, in this scenario, all the UEs are transmitting the same pilot symbols,
which leads to inter-UE interference at the BS.
When $T_{\tau} \ge \Nu$, the orthogonal pilots are assigned to difference antenna elements of one UE to avoid intra-UE interference.
When $T_{\tau} < \Nu$, the intra-UE interference is inevitable and, in the extreme case when $T_{\tau} = 1$, all the antenna elements
of UEs transmit the same pilot symbols, and therefore interfere at the BS.
For the sake of simplicity, we also assume that
all the UEs are located at the same distance from the BS and therefore experience the same path loss.

Fig.~\ref{fig: NMSE_Sc3} illustrates the average normalized MSE 
against the number of antenna elements at the \acp{UE} for a network with $K=2$ UEs.
This figure manifests similar behaviors as of Fig.~\ref{fig: NMSE_Sc2}, such as better channel estimate with higher $\Nu$, higher $\Nb$, and lower $\Np$. Moreover, comparing Fig.~\ref{fig: NMSE_Sc3_M=8} to Fig.~\ref{fig: NMSE_Sc3_M=1024} reveals that more antenna elements at the \ac{BS} increases separability of different \acp{UE} in the spatial domain, so makes it possible to use smaller number of unique pilots for a given power for the pilot contamination term.
In particular, with $\Nb = 8$, the
inter-UE interference term of the pilot contamination is so strong that additional interference
and the corresponding SINR loss
due to pilot reuse within one \ac{UE} (namely $\Tt < \Np$)
may not be tolerable and leads to substantial loss in the performance of channel estimation.
However, when $\Nb = 1024$, the inter-UE interference term is almost negligible, making intra-UE interference tolerable
for a given minimum SINR threshold for the received pilot signal.

\begin{figure}
 \centering
  {\footnotesize 
%
%
\definecolor{mycolor1}{rgb}{0.30588,0.39608,0.58039}%
\definecolor{mycolor2}{rgb}{0.85000,0.32500,0.09800}%
\definecolor{mycolor3}{rgb}{0.00000,0.74902,0.74902}%
\begin{tikzpicture}

\begin{axis}[%
width=0.85\columnwidth,
height=0.425\columnwidth,
at={(0\textwidth,0\textwidth)},
scale only axis,
xmin=1,
xmax=50,
xlabel={Number of UEs in the cell $K$},
ymin=-25,
ymax=0,
ylabel={Average normalized MSE (dB)},
axis background/.style={fill=white},
legend style={at={(0.97,0.03)},anchor=south east,legend cell align=left,align=left,draw=white!15!black},
ytick={-30,-25,-20,-15,-10,-5,0},
]
\addplot [color=mycolor1,solid,line width=1.0pt,mark size=1.5pt,mark=*,mark options={solid}]
  table[row sep=crcr]{%
1	-25.0430802660295\\
2	-14.9204566070479\\
3	-11.8871704369235\\
4	-10.4470915756302\\
6	-8.24899205489203\\
8	-6.85509168871749\\
10	-6.26732895228173\\
15	-5.26816046536708\\
20	-4.25942892835314\\
30	-3.28323879357054\\
40	-2.79231332344729\\
50	-2.43593664338991\\
};
\addlegendentry{$\Tt = 1$};

\addplot [color=mycolor2,line width=1.0pt,solid,mark=o,mark options={solid}]
  table[row sep=crcr]{%
1	-24.9338343830933\\
2	-24.1244493516087\\
3	-21.3363412263873\\
4	-20.7372078072949\\
6	-19.2408594550642\\
8	-18.2124977190251\\
10	-17.2788972550616\\
15	-14.2670118880573\\
20	-12.4411196046676\\
30	-10.190166168641\\
40	-8.83698909155421\\
50	-8.15230708382957\\
};
\addlegendentry{$\Tt = 4$};

\end{axis}
\end{tikzpicture}%
}
 \caption{Channel estimation performance in the precoded and combined pilot transmission scenario~($\sth$) as a function of number of UEs in the cell.}
\label{fig: NMSE_Sc3vsK}
\end{figure}
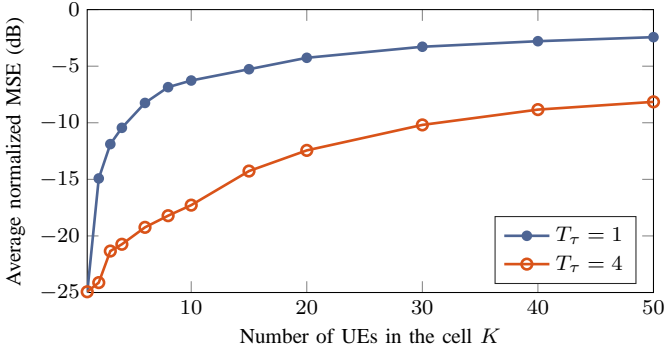

The pilot contamination effect on the channel estimation performance 
is investigated in Fig.~\ref{fig: NMSE_Sc3vsK}.
This figure illustrates the average normalized MSE against the number of UEs in the cell for a network with $\Nb = 128$, $\Nu = 32$ and $\Np = 4$.
For the sake of simplicity, all the UEs are located at the same distance from the BS and therefore experience the same path loss. The AoDs and AoAs
are drawn independently from uniform random distributions in $[-\pi/6,\pi/6]$ and $[-\pi/3,\pi/3]$, respectively.
The results are averaged over 50 different realizations of AoAs and AoDs.
Two pilot sequence length, namely $\Tt = 4,1$, are considered in this figure.
In the case of $\Tt = 4$, no intra-UE interference exists, and only the inter-UE interference
degrades the channel estimation performance when $K$ increases.
However, when $\Tt = 1$ both intra-UE and inter-UE interference
contaminate the pilot symbols which leads to at least 5 dB poorer performance compared to the case when $\Tt = 4$.

Fig.~\ref{fig: NMSE_Sc3_2} shows the average normalized MSE against the total pilot transmission power using the same setup as in Fig.~\ref{fig: NMSE_Sc3vsK}
when there are $K=2$ UEs in the network.
According to this figure, pilot precoding with $\Nu = 32$ can achieve the same estimation error as $\so$ with almost 5~dB less power.
Alternatively, for a given total pilot transmission power, pilot precoding can improve the estimation
error by 6~dB compared with the conventional $\so$.
Moreover, the performance of $\sth$ with only 4 orthogonal pilots is almost identical to that of $\st$ with 64 orthogonal pilots.
In other words, these 16 x shorter training length allows channel estimation of substantially higher number of \acp{UE} within the same coherence budget,
while also leaving more time for the data transmission.
Therefore, in $\sth$
we expect a boost in the achievable rate with 4 orthogonal pilots compared to that of $\st$.

\begin{figure}[t]
 \centering
  {\footnotesize 
%
%
\definecolor{mycolor1}{rgb}{0.30588,0.39608,0.58039}%
\definecolor{mycolor2}{rgb}{0.85000,0.32500,0.09800}%
\definecolor{mycolor3}{rgb}{0.00000,0.74902,0.74902}%
\begin{tikzpicture}

\begin{axis}[%
width=0.85\columnwidth,
height=0.425\columnwidth,
at={(0\textwidth,0\textwidth)},
scale only axis,
xmin=-20,
xmax=20,
xlabel={Pilot transmition energy $\PilotEnergy$ (dB)},
ymin=-50,
ymax=0,
ylabel={Average normalized MSE (dB)},
axis background/.style={fill=white},
legend style={at={(0.01,0.01)},anchor=south west,legend cell align=left,align=left,draw=white!15!black},
ytick={-50,-40,-30,-20,-10,0}
]
\addplot [color=mycolor1,solid,line width=1.0pt,mark=o,mark options={solid}]
  table[row sep=crcr]{%
-30	-0.132675056261493\\
-25 	-0.351362841966601\\
-20 	-1.197682976826604\\
-15 	-3.040719647757064\\
-10	-6.47480051403672\\
-5		-10.714655593915\\
0		-15.4354407921847\\
5		-20.3424104803166\\
10		-25.3132892192706\\
15		-30.3041221638664\\
20		-35.3012423175386\\
25		-40.3002245028422\\
30		-45.2999255360501\\
};
\addlegendentry{$\so$, $\Tt$=$64$};

\addplot [color=mycolor2,solid,line width=1.0pt,mark size =2.5pt,mark=+,mark options={solid}]
  table[row sep=crcr]{%
-30	-1.277885860354022\\
-25	-3.092017354599726\\
-20	-6.326552289904299\\
-15	-10.544903685921332\\
-10	-15.4299411742539\\
-5	-20.3353425637194\\
0	-25.3035801190752\\
5	-30.2935854896474\\
10	-35.2906254200734\\
15	-40.2894582781663\\
20	-45.2891410293164\\
25	-50.2890801499844\\
30	-55.2890287639658\\
};
\addlegendentry{$\st$, $\Tt$=$8$};

\addplot [color=mycolor3,solid,line width=1.0pt,mark size =1.5pt, mark=*,mark options={solid}]
  table[row sep=crcr]{%
-30	-1.289463400975208
-25	-3.101700984111667\\
-20	-6.308211054687329\\
-15	-10.455462411910844\\
-10	-15.2348338555142\\
-5	-19.9473832005295\\
0	-24.5480627566293\\
5	-29.0584674932908\\
10	-33.497035270826\\
15	-37.9612358372075\\
20	-42.630128557713\\
25	-47.4435725570186\\
30	-52.3643155751982\\
};
\addlegendentry{$\sth$, $\Tt$=$4$};

\addplot [color=mycolor3,dashdotted,line width=1.0pt,mark size =1.5pt,mark=*,mark options={solid}]
  table[row sep=crcr]{%
-30	-1.307986904320120\\
-25	-3.107118126507733\\
-20	-6.152008164297939\\
-15	-9.984527155293462\\
-10	-14.1084707646913\\
-5	-17.5918043352038\\
0	-20.2933252475602\\
5	-22.1996202411892\\
10	-23.2677667136668\\
15	-24.2709583024931\\
20	-25.0587994131842\\
25	-26.809761449675\\
30	-28.6409082434529\\
};
\addlegendentry{$\sth$, $\Tt$=$2$};

\addplot [color=mycolor3,dashed,line width=1.0pt,mark size =1.5pt,mark=*,mark options={solid}]
  table[row sep=crcr]{%
-30	-1.461658989234515\\
-25	-3.258790191403453\\
-20	-6.215927122899294\\
-15	-9.617642722650873\\
-10	-12.680036200332\\
-5	-14.2698710524526\\
0	-15.2687667738733\\
5	-15.682308765562\\
10	-15.6106698490122\\
15	-15.5025466890048\\
20	-15.5703670747657\\
25	-15.6707405200663\\
30	-15.4176686215269\\
};
\addlegendentry{$\sth$, $\Tt$=$1$};

\end{axis}
\end{tikzpicture}%
}
 \caption{The impact of pilot energy and number of transmitted pilot symbols on the channel estimation performance in the
three pilot transmission scenarios.}
\label{fig: NMSE_Sc3_2}
\end{figure}
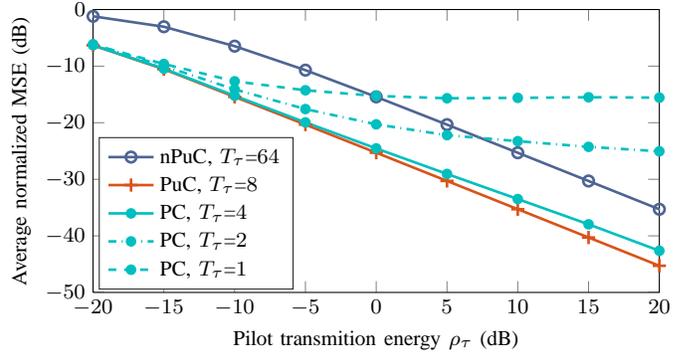


So far, we have analyzed the channel estimation quality of three pilot transmission scenarios. In the next section, we investigate their effects on the achievable data rate.

\section{Data Transmission}
\label{sec:Data Transmission}
In this section, we study the sum-rate of the multiple access channel between the UEs and the BS in the uplink data transmission phase.
Denoting the transmitted signal vector of UE $k$ by $\mathbf{x}_k = \mathbf{F}_k \mathbf{s}_k$,
\eqref{eq: data transmission} can be equivalently written as
\be \label{eq: data transmission_compact}
\begin{split}
\mathbf{y}_{d} =  \sum_{k=1}^{K} \mathbf{H}_k\mathbf{x}_k + \mathbf{z}_d = \sum_{k=1}^{K}\Hh_k \mathbf{x}_k +\underbrace{\sum_{k=1}^{K}\Ht_k \mathbf{x}_k  + \mathbf{z}_d}_{\mathbf{z}_{\text{eff}}} \:,
\end{split}
\ee
where $\mathbf{z}_{\text{eff}}$ is the effective noise at the BS which combines the receiver noise and the residual channel estimation errors.


In the following, we first present
the performance metric used for the performance evaluation of the network.
Subsequently, we study the precoding methods that are used in the data
transmission phase. 

\subsection{Performance Metric}
Although the optimal distribution of $\{\mathbf{x}_k\}_{k=1}^{K}$ is not known, using
\eqref{eq: data transmission_compact},
the following lower bound for the sum-rate of the network can be found as~\cite{Yoo2006}:
\be \label{eq: sum-rate}
\begin{split}
r &= \frac{\Td}{\Tc}\E\left[\log\det \left(\I_{\Nb} + \mathbf{R}_{\mathbf{z}_{\text{eff}}}^{-1}\sum_{k=1}^{K}\Hh_k \mathbf{R}_{\mathbf{x}_k} \Hh_k\herm \right)\right],
\end{split}
\ee
where $\mathbf{R}_{\mathbf{x}_k}$ and $\mathbf{R}_{\mathbf{z}_{\text{eff}}}$ are the covariance matrices of $\mathbf{x}_k$ and $\mathbf{z}_{\text{eff}}$, respectively, and are defined as

\begin{align}
\label{eq: covariance of X}
\mathbf{R}_{\mathbf{x}_k} &= \E\left[\mathbf{x}_k\mathbf{x}_k\herm\right]
= \mathbf{F}_k\E\left[\mathbf{s}_k\mathbf{s}_k\herm\right] \mathbf{F}_k\herm
= \frac{\DataEnergy}{\Td}\mathbf{F}_k\mathbf{F}_k\herm \:, \\
\label{eq: covariance of Z_eff}
\mathbf{R}_{\mathbf{z}_{\text{eff}}} &= \E\left[\mathbf{z}_{\text{eff}} \mathbf{z}_{\text{eff}}\herm\right]
= \sum_{k=1}^{K} \E\left[\Ht_k\mathbf{R}_{\mathbf{x}_k}\Ht_k\herm\right] +  \NoisePower\; \I_{\Nb}.
\end{align}

\subsection{Data Precoding}
For a single link MIMO network, when only imperfect CSI is available at the transmitter,
it is shown that the eigenvectors of the channel estimate covariance matrix represent the optimal transmit directions~\cite[Theorem 1]{Wang2009}.
More specifically,
the precoder of UE $k$ is designed in a way that its signal is transmitted in
the direction of the right singular vectors of the estimated channel matrix $\Hh_k$.
For the sake of simplicity,
we choose to allocate equal powers to different eigendirections for all three pilot precoding scenarios.
Note that
considering optimal power allocation does not change the insights
that we gain from the comparative performance analysis of this paper, but
complicates the mathematical analysis; see \cite{Wang2009} for the optimal power allocation algorithm.
Formally, denote the eigen-value decomposition of $\mathbf{R}_{\mathbf{x}_k}$ by $\mathbf{R}_{\mathbf{x}_k} = \bar{\mathbf{F}}_k \mathbf{\Lambda}_k\bar{\mathbf{F}}_k\herm$
and that of $\Hh_k\herm\Hh_k$ by $\Hh_k\herm\Hh_k = \mathbf{E}_k\mathbf{\Sigma}_k\mathbf{E}_k\herm$.
Here, we set $\bar{\mathbf{F}}_k = \mathbf{E}_k$ and
$\mathbf{\Lambda}_k = \DataEnergy/(\Td\Np)\mathbf{T}_{\Np}$ where $\mathbf{T}_{\Np}$ is a diagonal matrix
whose first $\Np$ diagonal elements are one and the rest are zero.
Comparing the designed covariance matrix of the data signal with the one in \eqref{eq: covariance of X}, the data precoding matrix corresponding to UE $k$ is found as
\be \label{eq: data precoder}
\mathbf{F}_k = \sqrt{\frac{1}{\Np}} \; \bar{\mathbf{F}}_k\mathbf{T}_{\Np} = \sqrt{\frac{1}{\Np}} \; \mathbf{E}_k\mathbf{T}_{\Np} \:.
\ee

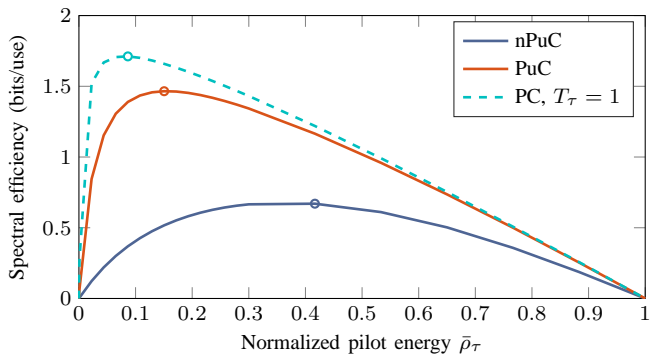
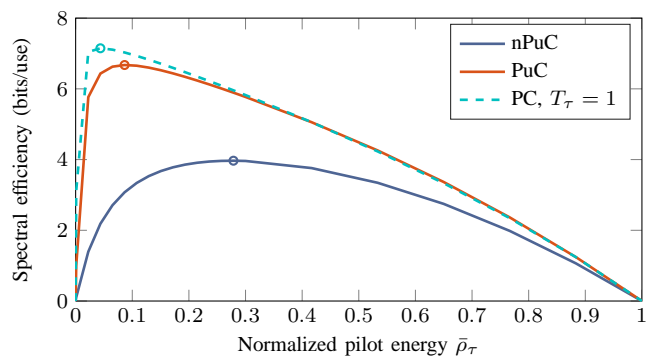
\begin{figure}[t]
\begin{subfigure}[t]{\columnwidth}
 \centering
  {\footnotesize 
%
%
\definecolor{mycolor1}{rgb}{0.30588,0.39608,0.58039}%
\definecolor{mycolor2}{rgb}{0.85000,0.32500,0.09800}%
\definecolor{mycolor3}{rgb}{0.00000,0.74902,0.74902}%
\begin{tikzpicture}

\begin{axis}[%
width=0.85\columnwidth,
height=0.425\columnwidth,
at={(0\textwidth,0\textwidth)},
scale only axis,
xmin=0,
xmax=1,
ymin=0,
ymax=2,
ylabel={Spectral efficiency (bits/use)},
xlabel={Normalized pilot energy $\bar{\rho}_{\tau}$},
axis background/.style={fill=white},
legend style={legend cell align=left,align=left,draw=white!15!black},
xtick = {0,0.1,0.2,0.3,0.4,0.5,0.6,0.7,0.8,0.9,1}
]
\addplot [color=mycolor1,solid,line width=1.0pt]
  table[row sep=crcr]{%
0	0\\
0.001	0.00589185015308965\\
0.0223571428571429	0.122746344763532\\
0.0437142857142857	0.21954987174333\\
0.0650714285714286	0.300526266177268\\
0.0864285714285714	0.368740989601247\\
0.107785714285714	0.426471004885234\\
0.129142857142857	0.475441537933831\\
0.1505	0.516982473953836\\
0.171857142857143	0.552134529017689\\
0.193214285714286	0.581723076462786\\
0.214571428571429	0.606410586117223\\
0.235928571428571	0.626734587913945\\
0.257285714285714	0.64313563167633\\
0.278642857142857	0.655978202844175\\
0.3	0.665566593528989\\
0.4167	0.670095768893336\\
0.53335	0.610708785598478\\
0.65	0.502657192302735\\
0.76665	0.357616859741892\\
0.8833	0.186998193866388\\
1	0\\
};
\addlegendentry{$\so$};

\addplot [color=mycolor2,solid,line width=1.0pt]
  table[row sep=crcr]{%
0	0\\
0.001	0.0664035444653485\\
0.0223571428571429	0.844904782175536\\
0.0437142857142857	1.15313764076512\\
0.0650714285714286	1.30637662331933\\
0.0864285714285714	1.38944675049001\\
0.107785714285714	1.43475030534566\\
0.129142857142857	1.45734476424228\\
0.1505	1.46517896314811\\
0.171857142857143	1.46284543067314\\
0.193214285714286	1.453182865093\\
0.214571428571429	1.43804051978078\\
0.235928571428571	1.41867486961646\\
0.257285714285714	1.39596961912235\\
0.278642857142857	1.37056449869235\\
0.3	1.34293411505807\\
0.4167	1.16509851465912\\
0.53335	0.959173356136063\\
0.65	0.73594319480405\\
0.76665	0.500044366112424\\
0.8833	0.25387965369218\\
1	0\\
};
\addlegendentry{$\st$};


\addplot [color=mycolor3,dashed,line width=1.0pt]
  table[row sep=crcr]{%
0	0\\
0.001	0.281999746404988\\
0.0223571428571429	1.52076921578525\\
0.0437142857142857	1.66805241845797\\
0.0650714285714286	1.70661983371895\\
0.0864285714285714	1.71137584939533\\
0.107785714285714	1.70102011361668\\
0.129142857142857	1.68244352498748\\
0.1505	1.65878356912684\\
0.171857142857143	1.63167746346394\\
0.193214285714286	1.60206632095221\\
0.214571428571429	1.57053221954601\\
0.235928571428571	1.53745676216594\\
0.257285714285714	1.50310247079742\\
0.278642857142857	1.46765753687405\\
0.3	1.43126182779915\\
0.4167	1.21940494059938\\
0.53335	0.991687844870363\\
0.65	0.75308005525213\\
0.76665	0.507161698750139\\
0.8833	0.256945016357152\\
1	0\\
};
\addlegendentry{$\sth$, $\Tt = 1$};

\addplot [color=mycolor1,only marks,mark size=1.5pt,mark=o,mark options={thick,solid}]
  table[row sep=crcr]{%
0.4167	0.670095768893336\\
};

\addplot [color=mycolor2,only marks,mark size=1.5pt,mark=o,mark options={thick,solid}]
  table[row sep=crcr]{%
0.1505	1.46517896314811\\
};


\addplot [color=mycolor3,only marks,mark size=1.5pt,mark=o,mark options={thick,solid}]
  table[row sep=crcr]{%
0.0864285714285714	1.71137584939533\\
};

\end{axis}
\end{tikzpicture}%
}
 \caption{Total energy $\PilotEnergy+\DataEnergy=0.1$.}
\label{fig: pilot-data tradeoff_rho=0.1}
\hfill
\end{subfigure}
\begin{subfigure}[t]{\columnwidth}
 \centering
  {\footnotesize 
%
%
\definecolor{mycolor1}{rgb}{0.30588,0.39608,0.58039}%
\definecolor{mycolor2}{rgb}{0.85000,0.32500,0.09800}%
\definecolor{mycolor3}{rgb}{0.00000,0.74902,0.74902}%
\begin{tikzpicture}

\begin{axis}[%
width=0.85\columnwidth,
height=0.425\columnwidth,
at={(0\textwidth,0\textwidth)},
scale only axis,
xmin=0,
xmax=1,
ymin=0,
ymax=8,
ylabel shift={4pt},
ylabel={Spectral efficiency (bits/use)},
xlabel={Normalized pilot energy $\bar{\rho}_{\tau}$},
axis background/.style={fill=white},
legend style={legend cell align=left,align=left,draw=white!15!black},
xtick = {0,0.1,0.2,0.3,0.4,0.5,0.6,0.7,0.8,0.9,1}
]
\addplot [color=mycolor1,solid,line width=1.0pt]
  table[row sep=crcr]{%
0	0\\
0.001	0.102881805587576\\
0.0223571428571429	1.39500052107175\\
0.0437142857142857	2.18512719360702\\
0.0650714285714286	2.7091874403642\\
0.0864285714285714	3.07492641064874\\
0.107785714285714	3.33850948820724\\
0.129142857142857	3.5320547865315\\
0.1505	3.67523052051097\\
0.171857142857143	3.78073074651654\\
0.193214285714286	3.85709458302949\\
0.214571428571429	3.91026013282616\\
0.235928571428571	3.94446963107\\
0.257285714285714	3.96282137543708\\
0.278642857142857	3.96761952644452\\
0.3	3.96060326646176\\
0.4167	3.75940222550332\\
0.53335	3.34365266009935\\
0.65	2.74672880805807\\
0.76665	1.98238985017576\\
0.8833	1.06531292202516\\
1	0\\
};\addlegendentry{$\so$};
\addplot  [color=mycolor2,solid,line width=1.0pt]
  table[row sep=crcr]{%
0	0\\
0.001	0.993105938197107\\
0.0223571428571429	5.76570842267866\\
0.0437142857142857	6.43205673849309\\
0.0650714285714286	6.62731419442631\\
0.0864285714285714	6.67386218810293\\
0.107785714285714	6.65466454103708\\
0.129142857142857	6.60062508867843\\
0.1505	6.52581759634527\\
0.171857142857143	6.43751953072205\\
0.193214285714286	6.3398337770689\\
0.214571428571429	6.23521850871968\\
0.235928571428571	6.12521089739368\\
0.257285714285714	6.01080017162719\\
0.278642857142857	5.89263337083087\\
0.3	5.77113520686405\\
0.4167	5.0563611746647\\
0.53335	4.2580248177982\\
0.65	3.36546336156744\\
0.76665	2.36440035942154\\
0.8833	1.24307461469267\\
1	0\\
};\addlegendentry{$\st$};
\addplot  [color=mycolor3,dashed,line width=1.0pt]
  table[row sep=crcr]{%
0	0\\
0.001	3.11511040835864\\
0.0223571428571429	7.04394134871339\\
0.0437142857142857	7.14741838435325\\
0.0650714285714286	7.10823100046832\\
0.0864285714285714	7.02890542502576\\
0.107785714285714	6.93186670022834\\
0.129142857142857	6.82492710075844\\
0.1505	6.711545128305\\
0.171857142857143	6.59349944585966\\
0.193214285714286	6.47180314735853\\
0.214571428571429	6.34707615452018\\
0.235928571428571	6.21971698312893\\
0.257285714285714	6.08998977556749\\
0.278642857142857	5.95807179857061\\
0.3	5.82408098766123\\
0.4167	5.05728466264029\\
0.53335	4.23020129604206\\
0.65	3.33156677735292\\
0.76665	2.34347277382453\\
0.8833	1.24212197718373\\
1	0\\
};\addlegendentry{$\sth$, $\Tt = 1$};
\addplot[color=mycolor1,only marks,mark size=1.5pt,mark=o,mark options={thick,solid}]
  table[row sep=crcr]{%
0.278642857142857	3.96761952644452\\
};
\addplot [color=mycolor2,only marks,mark size=1.5pt,mark=o,mark options={thick,solid}]
  table[row sep=crcr]{%
0.0864285714285714	6.67386218810293\\
};
\addplot [color=mycolor3,only marks,mark size=1.5pt,mark=o,mark options={thick,solid}]
  table[row sep=crcr]{%
0.0437142857142857	7.14741838435325\\
};
\end{axis}
\end{tikzpicture}%
}
 \caption{Total energy $\PilotEnergy+\DataEnergy=0.5$.}
 \label{fig: pilot-data tradeoff_rho=0.5}
\end{subfigure}
 \caption{Spectral efficiency as a function of normalized pilot energy $\bar{\rho}_{\tau} = \PilotEnergy/(\PilotEnergy+\DataEnergy)$, $\Nb = 128$ and $\Nu = 32$. The maximum of the curves are marked by circles.}
\label{fig: pilot-data tradeoff}
\end{figure}

Figs.~\ref{fig: pilot-data tradeoff} and~\ref{fig: impact of M and N on rho and r} show the spectral efficiency of a network employing the aforementioned data precoding method. In these simulations, we assume that the channel estimation is performed according to the scenarios presented in Section~\ref{sec: Channel Estimation}. We consider a cell with $K=2$ UEs located at the same distance from the BS, where the AoAs and AoDs are set similarly to the ones in Fig.~\ref{fig: NMSE_Sc1}.
We also assume that each coherence block has length $\Tc = 128$ channel uses and the number of paths between each UE and the BS is $\Np = 4$.

As we increase the pilot transmission energy $\PilotEnergy$, the channel estimation performance improves in all pilot transmission scenarios; see Fig.~\ref{fig: NMSE_Sc3_2}.
Acquiring more accurate CSI, better precoders can be designed for the data transmission phase.
However, since the total energy is fixed, increasing the pilot energy leaves less energy for the data transmission, leading to a well-known trade-off between the amount of energy allocated to pilot and data transmission~\cite{HassibiHochwald2003}.
This trade-off is illustrated in Fig.~\ref{fig: pilot-data tradeoff} for three pilot transmission scenarios.
From both Figs.~\ref{fig: pilot-data tradeoff_rho=0.1} and~\ref{fig: pilot-data tradeoff_rho=0.5},
$\sth$ and $\st$ always lead to higher spectral efficiencies than the conventional $\so$.
Moreover, the maximum spectral efficiency that $\sth$ can reach is higher than $\st$, though $\st$ has a better channel estimation performance; see Fig.~\ref{fig: NMSE_Sc3_2}.
Another observation from Fig.~\ref{fig: pilot-data tradeoff} is that the optimal pilot energy of $\sth$ and $\st$ are substantially smaller than that of $\so$. Indeed, by pilot precoding and combining, a minimal amount of pilot power that is allocated to a few (possibly one) pilot symbols $T_{\tau} = 1$ may lead to a very high spectral efficiency.
Moreover, increasing the total energy shifts these optimal points to the left, implying that a smaller fraction of the total energy needs to be allocated to the pilot transmission phase.
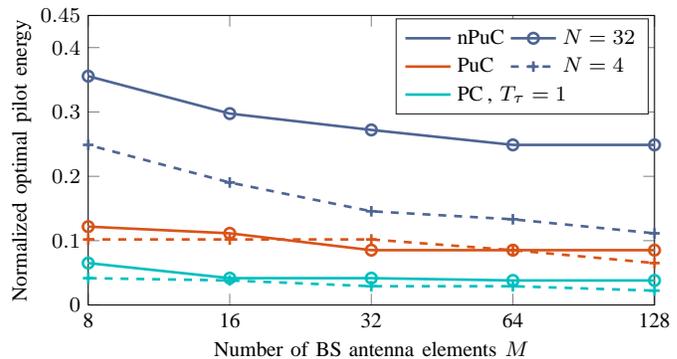
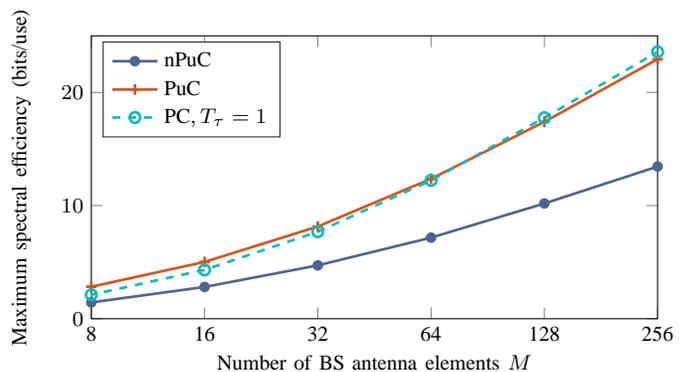
\begin{figure}
\begin{subfigure}[t]{\columnwidth}
 \centering
  {\footnotesize 
%
%
\definecolor{mycolor1}{rgb}{0.30588,0.39608,0.58039}%
\definecolor{mycolor2}{rgb}{0.85000,0.32500,0.09800}%
\definecolor{mycolor3}{rgb}{0.00000,0.74902,0.74902}%
\begin{tikzpicture}[trim axis right]

\begin{axis}[%
width=0.85\columnwidth,
height=0.435\columnwidth,
at={(0\textwidth,0\textwidth)},
scale only axis,
xmode=log,
xmin=8,
xmax=128,
xminorticks=true,
xlabel={Number of BS antenna elements $\Nb$},
ylabel shift={-2pt},
ymin=0,
ymax=0.45,
ylabel={Normalized optimal pilot energy},
axis background/.style={fill=white},
legend style={at={(0.99,0.99)},anchor=north east,legend cell align=left,align=left,draw=white!15!black,legend columns=2},
ytick={0,0.1,0.2,0.3,0.4,0.45},
xtick={8,16,32,64,128},
xticklabels = {8,16,32,64,128}
]
\addplot [color=mycolor1,solid,line width=1.0pt]
  table[row sep=crcr]{%
4	0.174107037757454\\
};\addlegendentry{$\so$};
\addplot [color=mycolor1,solid,line width=1.0pt,mark=o,mark options={solid}]
  table[row sep=crcr]{%
4	0.174107037757454\\
};\addlegendentry{$\Nu=32$};

\addplot [color=mycolor2,solid,line width=1.0pt]
  table[row sep=crcr]{%
4	0.145634847750124\\
};\addlegendentry{$\st$};

\addplot [color=mycolor1,dashed,line width=1.0pt,mark=+,mark options={solid}]
  table[row sep=crcr]{%
4	0.174107037757454\\
};\addlegendentry{$\Nu=4$};

\addplot [color=mycolor3,solid,line width=1.0pt]
  table[row sep=crcr]{%
4	0.145634847750124\\
};\addlegendentry{$\sth$};
\addplot [color=white,solid]
  table[row sep=crcr]{%
4	0.145634847750124\\
};\addlegendentry{\hspace{-10mm}, $\Tt=1$};

\addplot [color=mycolor1,dashed,line width=1.0pt,mark=+,mark options={solid}]
  table[row sep=crcr]{%
4	0.174107037757454\\
8	0.248838954136854\\
16	0.190367074884342\\
32	0.145634847750124\\
64	0.133195574662462\\
128	0.111413745744046\\
};

\addplot [color=mycolor2,dashed,line width=1.0pt,mark=+,mark options={solid}]
  table[row sep=crcr]{%
4	0.145634847750124\\
8	0.101897438140199\\
16	0.101897438140199\\
32	0.101897438140199\\
64	0.0852338772792677\\
128	0.0652057229342861\\
};


\addplot [color=mycolor3,dashed,line width=1.0pt,mark=+,mark options={solid}]
  table[row sep=crcr]{%
4	0.0223346726631486\\
8	0.041726135425828\\
16	0.0381621340794936\\
32	0.0291948415442296\\
64	0.0291948415442296\\
128	0.0223346726631486\\
};

\addplot [color=mycolor1,solid,line width=1.0pt,mark=o,mark options={solid}]
  table[row sep=crcr]{%
4	0.355648030622313\\
8	0.355648030622313\\
16	0.297487956026606\\
32	0.272078282569504\\
64	0.248838954136854\\
128	0.248838954136854\\
};

\addplot [color=mycolor2,solid,line width=1.0pt,mark=o,mark options={solid}]
  table[row sep=crcr]{%
4	0.145634847750124\\
8	0.121818791201012\\
16	0.111413745744046\\
32	0.0852338772792677\\
64	0.0852338772792677\\
128	0.0852338772792677\\
};


\addplot [color=mycolor3,solid,line width=1.0pt,mark=o,mark options={solid}]
  table[row sep=crcr]{%
4	0.0498837602969998\\
8	0.0652057229342861\\
16	0.041726135425828\\
32	0.041726135425828\\
64	0.0381621340794936\\
128	0.0381621340794936\\
};

\end{axis}
\end{tikzpicture}%
}
 \caption{\label{fig: impact of M and N on rho}}
\hfill
\end{subfigure}
\begin{subfigure}[t]{\columnwidth}
 \centering
  {\footnotesize 
%
%
\definecolor{mycolor1}{rgb}{0.30588,0.39608,0.58039}%
\definecolor{mycolor2}{rgb}{0.85000,0.32500,0.09800}%
\definecolor{mycolor3}{rgb}{0.00000,0.74902,0.74902}%
\begin{tikzpicture}

\begin{axis}[%
width=0.85\columnwidth,
height=0.425\columnwidth,
at={(0\textwidth,0\textwidth)},
scale only axis,
xmode=log,
ylabel shift={6pt},
xmin=8,
xmax=256,
xminorticks=true,
xlabel={Number of BS antenna elements $\Nb$},
ymin=0,
ymax=25,
ylabel={Maximum spectral efficiency (bits/use)},
axis background/.style={fill=white},
legend style={at={(0.02,0.98)},anchor=north west,legend cell align=left,align=left,draw=white!15!black},
xtick={8,16,32,64,128,256},
xticklabels = {8,16,32,64,128,256}
]

\addplot [color=mycolor1,solid,line width=1.0pt,mark size=1.5pt,mark=*,mark options={solid}]
  table[row sep=crcr]{%
4	0.717717987737343\\
8	1.44138624156558\\
16	2.80753928993106\\
32	4.71596081075566\\
64	7.16628526424009\\
128	10.1832111689177\\
256	13.451364385344633\\
};\addlegendentry{$\so$};
\addplot [color=mycolor2,solid,line width=1.0pt,mark=+,mark options={solid}]
  table[row sep=crcr]{%
4	1.5561252225457\\
8	2.82855840145537\\
16	5.01973321255768\\
32	8.14922783196835\\
64	12.3657839187741\\
128	17.3978001083964\\
256 22.932322789515176\\
};\addlegendentry{$\st$};
\addplot [color=mycolor3,dashed,line width=1.0pt,mark=o,mark options={solid}]
  table[row sep=crcr]{%
4	0.975444737497424\\
8	2.12164748689817\\
16	4.31590503885746\\
32	7.6878620549646\\
64	12.2247977154297\\
128	17.7594668164763\\
256 23.580519140699707\\
};\addlegendentry{$\sth, \Tt=1$};
\end{axis}
\end{tikzpicture}%
}
 \caption{\label{fig: impact of M on r}}
\end{subfigure}
 \caption{Impact of number of antenna elements on the optimal pilot energy and maximum spectral efficiency when $\PilotEnergy + \DataEnergy = 4$.}
\label{fig: impact of M and N on rho and r}
\end{figure}

Fig.~\ref{fig: impact of M and N on rho and r} shows the optimal pilot energy and the corresponding maximum spectral efficiency against the number of antenna elements. From Fig.~\ref{fig: impact of M and N on rho}, the optimal pilot energy decreases with $\Nb$ in all three scenarios. On the contrary, the UEs with larger number of antennas need more pilot transmission energy to reach the optimal performance (this can be observed by comparing solid lines to the dashed ones). This increment is much higher in $\so$ than other schemes. In general, the optimal pilot energy in $\st$ and $\sth$ is less sensitive to the changes of $\Nb$ or of $\Nu$.
Fig.~\ref{fig: impact of M on r} presents the maximum spectral efficiency against $\Nb$ when $\Nu = 32$.
From this figure, employing either $\st$ or $\sth$ for pilot transmission significantly improves the maximum spectral efficiency compared to the conventional $\so$. In particular, the improvement is around 80\% with only 128 BS antennas.

The following proposition gives a lower bound for the achievable sum-rate using the aforementioned data precoding method and different pilot transmission scenarios of Section~\ref{sec: Channel Estimation}.
\begin{prop} \label{prop: rate in large antenna regime}
Consider the pilot transmission scenario $\sx \in \{\so,\st,\sth\}$ and uplink data transmission using the data precoding filter of \eqref{eq: data precoder} for UE $k$.
Assume that $\{\Au_k\}_{k=1}^{K}$ and $\{\Ab_k\}_{k=1}^{K}$ are perfectly known. If $\Nb,\Nu\to\infty$, then
\be \label{eq: lower bound for sum-rate}
r \ge \frac{\Np \Td}{\Tc}\log\left(1+\frac{\frac{\DataEnergy}{\Td }\sum_{k=1}^{K}\CG_k}{\Np\NoisePower}\right) \:.
\ee
\end{prop}
\begin{IEEEproof}
A proof is given in Appendix~B.
\end{IEEEproof}

\begin{corollary}
\label{corollary: maximum sum-rate}
For given budgets of coherence time $\Tc$ and total energy $\PilotEnergy + \DataEnergy$, the lower bound in \eqref{eq: lower bound for sum-rate} is maximized when $\PilotEnergy\to 0$ and $\Td/\Tc~\to~1$ (or equivalently $\Tt = 1$).
\end{corollary}
Corollary~\ref{corollary: maximum sum-rate} implies that in the large antenna regime, among the three pilot transmission scenarios, only $\sth$ can reach the maximum spectral efficiency.

\section{Further Discussions}
\label{sec: FurtherDiscussions}

Equation~\eqref{eq: antenna_config} implies that the vectors $\vec{b}(\theta)$ or $\vec{u}(\theta)$ with different $\theta$ create an asymptotically orthonormal basis, which can be used as orthogonal spatial signatures. More interestingly, asymptotically, there are infinitely many such spatial signatures (realized by changing $\theta$).
This leads to an interesting and practically relevant consequence of pilot precoding and combining:
maintaining orthogonality of pilots
in the spatial domain instead of code domain becomes possible.
To appreciate this aspect, we recall that in \ac{MU-MIMO} systems with single antenna \acp{UE},
the orthogonality of pilot signals must be maintained in the code domain to avoid intra-cell pilot contamination effects. Therefore, there is an inherent trade-off between the number of symbols spent on constructing the pilot sequences and the number of symbols available for data transmission, as illustrated by Fig.~\ref{fig: pilot-data tradeoff}.
Due to the proposed pilot precoding schemes, this trade-off can be relaxed by creating pilot signal separability in the spatial domain.
We have the following conceptually important result:
\begin{corollary}\label{corollary: asymptotic-zero-product-aBS}
Consider the channel model in~\eqref{eq:geometry-based channel model}.
Suppose that there are a limited number of paths between each UE and its serving BS and that the second-order statistics are perfectly known.
Let either the number of BS antennas $\Nb$ or the number of UE antennas $\Nu$ tends to infinity, then
1)~the pilot contamination problem disappears,
2)~a multi-cell network can be modeled by multiple uncoordinated single-cell networks with no performance loss\footnote{Note that inter-cell coordination may still bring gain to the resource allocation performance~\cite{shokri2016value}.}, and
3)~channel estimation in the entire network can be done with a single pilot symbol.
\end{corollary}
Corollary~\ref{corollary: asymptotic-zero-product-aBS} suggests that although we have considered a single-cell scenario throughout this paper, our insights are valid in a multicell scenario, especially in the large antenna regime.
In fact, all the interference components in the pilot transmission can be rejected either at the transmitter or at the receiver.
In particular, from~\eqref{eq: received signal-s3}, the transmitter cancels intra-UE interference, and the receiver cancels out inter-UE interference (which can be readily extended to the inter-cell interference).
Notably, $\sth$ allows using fixed-length pilot sequences (almost) independently of the number of \ac{MU-MIMO} users.
As an extreme case, a single pilot symbol can be used by multiple \acp{UE}.
In real life deployments, even though spatial separability based on channel covariance matrix knowledge is possible in $\sth$, multiple antenna \acp{UE} benefit from
code domain separability among pilot signals used to estimate the channel of
the different \ac{UE} antennas.
However, when the number of antennas at the \ac{BS} grows
large, full reuse of the same pilot sequence among the \ac{UE} antennas is possible.
This, in turn, maximizes the number of symbols available for data transmission.

\begin{table}[t]
  \centering
  \caption{Minimum number of unique pilots needed in the pilot transmission phase. ``UL'' and ``DL'' stand for uplink and downlink, respectively.}\label{table: pilots}
{
\renewcommand{\arraystretch}{1.4}
  {
   \begin{tabular}{|c|l|c|c|c|}
\cline{3-5}
\multicolumn{2}{c}{} \vline    									& $\so$				&$\st$					&$\sth$			\\ \hline
\multirow{4}{4em}{UL pilots}	&  $\Nu,\Nb < \infty $ 			
&
$K\Nu$	&
$K\Np$	&
$K\Np$	\\ \cline{2-5}
									&  $\Nu\to\infty,\Nb<\infty$	&
$K\Nu$	&
$K$		&  $K$								\\ \cline{2-5}
									&  $\Nu<\infty,\Nb\to\infty$	&
$K\Nu$ 	&	$K\Np$						&
$\Np$	\\ \cline{2-5}
									&  $\Nu\to\infty,\Nb\to\infty$	&
$K\Nu$	&
	$K$	&
$1$	\\ \hline \hline
\multirow{4}{4em}{DL pilots}	&  $\Nu,\Nb < \infty $ 			&   $\Nb$ 						&
$K\Np$ 	&
$K\Np$	\\ \cline{2-5}
									&  $\Nu\to\infty,\Nb<\infty$	& 	$\Nb$  						&	$K\Np$						&
$\Np$	\\ \cline{2-5}
									&  $\Nu<\infty,\Nb\to\infty$	&   $\Nb$ 						&
	$K$	&  $K$								\\ \cline{2-5}
									&  $\Nu\to\infty,\Nb\to\infty$	&   $\Nb$							&
$K$		&
 $1$	\\ \hline
\end{tabular}}
}
\end{table}
To generalize Corollary~\ref{corollary: asymptotic-zero-product-aBS}, Table~\ref{table: pilots} shows the number of unique pilots that the three pilot transmission scenarios need in different network settings.
This table also includes the case of downlink pilot transmission.
From this table, it is clear that the training sequence lengths required in $\sth$ and $\st$ are substantially smaller than that in $\so$.
The minimum number of unique pilots in both $\so$ and $\st$ scales up with the number of \acp{UE}; whereas there are situations in which $\sth$ may need a constant
number of pilots independently of the number of transmit antenna elements or the number of \acp{UE}.
Specifically, when the number of receive antennas ($\Nb$ in uplink and $\Nu$ in downlink) goes to infinity,
a maximum of $\Np$ pilot sequences is enough to handle the channel estimation of the entire network, irrespective of the number of \acp{UE} or \acp{BS}.
Also, note that large values of $\Nu$ can model wireless backhauling use cases.
Moreover, this table illustrates that, for example, as $\Nu \to \infty$ and $\Nb < \infty$), $\st$ may benefit
from uplink pilots whereas $\sth$ may benefit from downlink pilots.
This disagreement indicates that employing pilot precoding
and combining may relax the usual system design constraint
of always relying on the uplink pilots for CSI acquisition.



\section{Concluding Remarks and Outlook}
\label{Sec:Conc}

The performance of \ac{MU-MIMO}
systems heavily dependents on the
quality of the acquired CSI.
The UEs can facilitate the acquisition of high-quality CSI if they, contrarily to what commonly assumed, exploit the multi-antenna capabilities.
In this paper, we showed that UEs can precode pilot signals to
drastically improve the
quality of the \ac{CSI} at the BS, which in turn improves the spectral efficiency
during data transmission.
Moreover, the CSI further improves if pilot precoding is carried along with
pilot combining at the \ac{BS}.
The aforementioned gains are more prominent when the channels are sparse and the BS and UEs are equipped with a large number of antennas, both hold in
mmWave networks.


The insights of this work suggest a further development of the following major research questions

\begin{itemize}
\item
The \ac{MMSE} channel estimation used to obtain the estimated channel
in \eqref{eq: MMSE channel estimate s1} (which is then used throughout in Section \ref{sec: Channel Estimation})
inherently depends on exploiting side information lying in the second-order
statistics (covariance matrices) of the channel vectors.
As noted in \cite{Yin:13}, the role of covariance matrices is to capture structural information
related to the distribution of the multi-path AoA at the serving \ac{BS}.
Therefore, imperfect knowledge of the channel covariance matrices
entails channel estimation errors that result in precoding errors.
\item
Since pilot precoding increases the spatial separation of \acp{UE},
it potentially mitigates
the affects of pilot contamination by employing low rate multicell coordination techniques
proposed in, for example, \cite{Yin:13}.
This is because the method proposed in \cite{Yin:13} achieves better results if the
paths of different \acp{UE} do not overlap.
In the asymptotic regime,
if either the number of BS antennas $N_{\text{BS}}$ or the number of UE antennas $N_{\text{UE}}$ tends to infinity,
pilot contamination disappears since the \acp{UE} become perfectly orthogonal in the AoA domain.
Thereby, a multi-cell network can be modeled by multiple uncoordinated single-cell networks with no performance loss.
\item
Although in this paper we considered the case of UL pilot and data transmissions, the concept of precoded
pilots in  $\sth$ 
and $\st$ 
can be employed for DL pilot and data transmissions as well.
As noted earlier, DL pilot precoding can be useful in \ac{FDD} systems.
\end{itemize}

\section*{Appendix~A: Preliminaries}\label{appendix: preliminaries}
In this subsection, we give preliminary linear algebra to prove the results in Appendix~B.
\begin{defin}[Products~\cite{Liu2008}]
\label{def: Hadamard-Kronecker-Khatri}
The Hadamard product of any two arbitrary matrices $\mathbf{A}$ and $\mathbf{B}$, of the same size, is defined as
\be
\left[\mathbf{A} \circ \mathbf{B}\right]_{i,j} = \left[\mathbf{A}\right]_{i,j} \left[\mathbf{B}\right]_{i,j} \:.
\ee
The Kronecker product of $\mathbf{A}$ of size $M\times N$ and $\mathbf{B}$ of any arbitrary size is defined as
\be
\mathbf{A} \otimes \mathbf{B} =
\left[
\begin{matrix}
\left[\mathbf{A}\right]_{1,1} \mathbf{B} 	& \dots 	& \left[\mathbf{A}\right]_{1,N} \mathbf{B} \\
\vdots 											& \ddots 	& \vdots 											\\
\left[\mathbf{A}\right]_{M,1} \mathbf{B} 	& \dots 	& \left[\mathbf{A}\right]_{M,N} \mathbf{B}
\end{matrix}
\right]  \:.
\ee
The Khatri-Rao product of $\mathbf{A}$ of size $M\times L$ and
$\mathbf{B}$ of size $N\times L$ is defined as
\be
\mathbf{A} \odot \mathbf{B}=  \left[
\begin{matrix}
\left[\mathbf{A}\right]_{:,1} \otimes \left[\mathbf{B}\right]_{:,1} & \dots & \left[\mathbf{A}\right]_{:,L} \otimes \left[\mathbf{B}\right]_{:,L}
\end{matrix}
\right]  \:.
\ee
\end{defin}

\begin{lemma}[Vectorization lemma~\cite{Liu2008}]
\label{lemma: vectorization}
For any three matrices $\mathbf{A}$, $\mathbf{B}$, and $\mathbf{C}$ with appropriate dimensions, we have
\begin{align} \label{eq: vectorization}
\vect(\mathbf{A}\mathbf{B}\mathbf{C}) &= \left(\mathbf{C}\tran \otimes \mathbf{A} \right)\vect\left({\mathbf{B}}\right).
\end{align}
\end{lemma}

\begin{lemma}[\hspace{-0.05mm}\cite{Liu2008}]
\label{lemma: KhatriRao_kronecker}
Consider matrices $\mathbf{A}\in\mathbb{C}^{N\times M}$, $\mathbf{B}\in\mathbb{C}^{R\times S}$, $\mathbf{C}\in\mathbb{C}^{M\times L}$ and
$\mathbf{D}\in\mathbb{C}^{S \times L}$. The following equalities always hold:
\begin{align*}
\left(\mathbf{A}\otimes  \mathbf{B} \right) \left(\mathbf{C} \odot \mathbf{D}\right) &=
\left(\mathbf{A}\mathbf{C} \odot \mathbf{B}\mathbf{D}\right) \:,\\
\left(\mathbf{C} \odot \mathbf{D}\right)\herm \left(\mathbf{A} \otimes  \mathbf{B} \right) &=
\left(\mathbf{A}\mathbf{C} \odot \mathbf{B}\mathbf{D}\right)\herm \:, \\
\left(\mathbf{C} \odot \mathbf{D}\right)\herm\left(\mathbf{C} \odot \mathbf{D}\right) & =
\left(\mathbf{C}\herm \mathbf{C}\right)\circ\left(\mathbf{D}\herm \mathbf{D}\right) \:.
\end{align*}
\end{lemma}
\begin{lemma}
\label{lemma:tr_inv}
For any positive definite matrix $\mathbf{A} \in \mathbb{C}^{N\times N}$
\be
\sum_{i=1}^N \frac{1}{a_{ii}} \le \tr\left(\mathbf{A}^{-1}\right)
\le \frac{\big(1+ \cond\left(\mathbf{A}\right)\big)^2}{4\cond\left(\mathbf{A}\right)}
\sum\limits_{i=1}^{N}  \frac{1}{a_{ii}} \:,  \nonumber
\ee
where $a_{ii}$ and $\cond(\mathbf{A})$ are diagonal element $i$ and the condition number of matrix $\mathbf{A}$, respectively.
\end{lemma}
\begin{lemma}[\hspace{-0.05mm}\cite{Horn2012}]
\label{lemma: bound on trace of  product}
For any two \ac{PSD} matrices $\mathbf{A} \in \mathbb{C}^{N\times L}$ and $\mathbf{B} \in \mathbb{C}^{L\times M}$,
\be
\lambda_{\min}\left(\mathbf{A}\right) \tr\left(\mathbf{B}\right)\le \tr\left(\mathbf{A} \mathbf{B}\right) \le \lambda_{\max}\left(\mathbf{A}\right) \tr\left(\mathbf{B}\right) \:,  \nonumber
\ee
where $\lambda_{\min}\left(\mathbf{A}\right)$ and $\lambda_{\max}\left(\mathbf{A}\right)$ are the smallest and largest eigenvalues of $\mathbf{A}$, respectively.
\end{lemma}

\begin{lemma}
\label{lemma: inequality_for_delta}
For any two real values $0 < a \leq b$, we have
\be
\frac{1+a}{1+b} \geq \frac{a}{b} \:.
\ee
\end{lemma}

\section*{Appendix~B: Proofs}\label{appendix_proofs}
\subsection{Proof of Lemma~\ref{lemma:tr_inv}}
The upper bound is given in~\cite{Bai1996}, and the lower bound can be obtained by the Cauchy-Schwarz inequality. In particular, let $\mathbf{e}_i \in \mathbb{C}^{N}$ be the standard basis vector (with 1 in its $i$-th entry and 0 otherwise). Then,
\begin{align*}\label{eq:LB_proof}
1 = \left(\mathbf{e}_i^* \sqrt{\mathbf{A}}\sqrt{\mathbf{A}^{-1}} \mathbf{e}_i\right)^2 \,  &\stackrel{(a)}{\le} \,
\left(\mathbf{e}_i^* \mathbf{A} \mathbf{e}_i\right) \left(\mathbf{e}_i^* \mathbf{A}^{-1} \mathbf{e}_i\right) \nonumber \\
\frac{1}{\left[\mathbf{A}\right]_{ii}} & \, \le \left[\mathbf{A}^{-1}\right]_{ii} \; ,
\end{align*}
where (a) is due to the Cauchy-Schwarz inequality. The lower bound will be proved by summing both sides of
the second inequality over $i$.

\subsection{Proof of Proposition~\ref{prop: covariance of error_s1}}
From~\eqref{eq: pilot matrix_s1}, it is straightforward to show that
\be
\label{eq: PilotSignals_tilde}
\Pt_k^{(\so)} \left(\Pt_k^{(\so)}\right)\herm = \frac{\PilotEnergy}{\Nu}\I \; .
\ee
Now, the covariance matrix of $\widetilde{\mathbf{H}}_{k}^{(\so)}$ can be expressed as
\begin{align*}
\Rt_{k}^{(\so)} =& \E\left[\vect\left(\Ht_k^{(\so)}\right) \vect\left(\Ht_k^{(\so)}\right)\herm\right] \\
=& \mathbf{R}_{k} -
\mathbf{R}_{k} \Pt^{(\so)}_k \\
&\hspace{5mm} \cdot \left(\!\!\left(\!\Pt^{(\so)}_k\!\right)^{\mbox{\!\scriptsize H}}\! \mathbf{R}_{k} \Pt^{(\so)}_k \!+\! \NoisePower \I \!\right)^{\!\!\!-1\!\!}\!
\left(\!\Pt^{(\so)}_k\!\right)^{\mbox{\!\scriptsize H}}\!\mathbf{R}_{k} \\
\stackrel{(a)}{=}&\mathbf{R}_{k} - \mathbf{R}_{k} \left(\mathbf{R}_k + \frac{\Nu \NoisePower}{\PilotEnergy} \I \right)^{-1}\mathbf{R}_{k} \\
\stackrel{(b)}{=}& \left(\Au_k^* \odot \Ab_k \right) \bigg(\frac{\PilotEnergy}{\Nu\NoisePower}\left(\Au_k^* \odot \Ab_k \right)\herm\left(\Au_k^* \odot \Ab_k \right) \\
&\hspace{5mm}+ \frac{1}{\delta\Nb\CG_k}\I   \bigg)^{-1}
 \left(\Au_k^* \odot \Ab_k \right)\herm \:,
\end{align*}
where (a) is valid due to~\eqref{eq: PilotSignals_tilde} and (b) holds
by substituting for $\mathbf{R}_{k}$ from \eqref{eq: R_H} and applying matrix inversion lemma.
The proof will be completed using Lemma~\ref{lemma: KhatriRao_kronecker} and after some straightforward algebraic manipulations.

\subsection{Proof of Proposition~\ref{prop: bounds-on-estimation-error-S1}}
First, we note that
\be
\label{eq: trRH}
\begin{split}
\tr\left(\mathbf{R}_{k} \right) &= \tr\left( \delta \Np \sigma_k^2  \left(\Au_k^*\odot\Ab_k\right) \left(\Au_k^*\odot\Ab_k\right)\herm\right) \\
&\stackrel{(a)}{=}  \delta \Nb \sigma_k^2 \; \tr \left(\mathbf{R}_{\Au_k}\tran\circ\mathbf{R}_{\Ab_k}\right) \\
&\stackrel{(b)}{=} \delta \Nb \sigma_k^2 \Np\; ,
\end{split}
\ee
where (a) follows from the commutative property of trace and after applying Lemma~\ref{lemma: KhatriRao_kronecker} and (b) holds since $\left[\mathbf{R}_{\Au_k}\right]_{ii} = \left[ \mathbf{R}_{\Ab_k}\right]_{ii} = 1$ for $i = 1,\dots,\Np$.

Now, we are ready to prove Proposition~\ref{prop: bounds-on-estimation-error-S1}. Using Proposition~\ref{prop: covariance of error_s1},
the normalized channel estimation error is calculated as
\begin{alignat}{3}\label{eq:channel_estimation_error_sc1}
e_k^{(\so)} \!
& = {\tr\left(\Rt_k^{(\so)} \right)}/{\tr\left(\mathbf{R}_{k} \right)}  && \nonumber \\
& \stackrel{(c)}{=}\frac{1}{\Np}\tr\!\left(\!\left(\I_{\Np} \!+\!  \Nb\zeta_k \mathbf{R}_{\Au_k}\tran\!\circ\!\mathbf{R}_{\Ab_k}\! \right)^{\!\!-1}\!\!\left(\mathbf{R}_{\Au_k}\tran \!\circ\! \mathbf{R}_{\Ab_k} \!\right)\!\right)   \\
&=\!\frac{1}{\Nb \zeta_k} \!
\left(\!1\!-\! \frac{1}{\Np}
\tr\! \left(\! \left(\I_{\Np} \!+\!  \Nb\zeta_k \mathbf{R}_{\Au_k}\tran \!\!\circ\! \mathbf{R}_{\Ab_k}\!\right)^{\!\!-1}\right)\! \right) \;,  && \nonumber
\end{alignat}
where (c) follows after substituting for $\tr\left(\mathbf{R}_{k}\right)$ and $\Rt^{(\so)}_{k}$ from \eqref{eq: trRH} and \eqref{eq: covariance of error_s1}, respectively and applying Lemma~\ref{lemma: KhatriRao_kronecker}.
Considering that for $i = 1,\dots,\Np$
\begin{equation*}
\left[\I_{\Np} +  \Nb\zeta_k \mathbf{R}_{\Au_k}\tran \circ \mathbf{R}_{\Ab_k} \right]_{ii} =
1 + \Nb\zeta_k \:,
\end{equation*}
and applying Lemma~\ref{lemma:tr_inv} on the trace at the right hand side of~\eqref{eq:channel_estimation_error_sc1} along with the fact that $e_k^{(\so)}\ge 0$, the bounds in \eqref{eq: NMSE_s1} follows after some straightforward algebraic manipulations.

\subsection{Proof of Proposition~\ref{prop: covariance of error_s2}}
Similar to proof of Proposition~\ref{prop: covariance of error_s1}, we form the covariance matrix of $\Ht_k^{(\st)}$ as
\begin{align*}
\Rt_k^{(\st)} =& \E\left[\vect\left(\Ht_k^{(\st)}\right) \vect\left(\Ht_k^{(\st)}\right)\herm\right] \\
=& \mathbf{R}_{k}-\mathbf{R}_{k} \Pt^{(\st)}_k \\
&\hspace{5mm} \cdot \left(\!\left(\!\Pt^{(\st)}_k\!\right)^{\mbox{\!\scriptsize H}}\! \mathbf{R}_{k} \Pt^{(\st)}_k \!+\! \NoisePower \I \!\right)^{\!\!\!-1\!\!}\!
\left(\!\Pt^{(\st)}_k\!\right)^{\mbox{\!\scriptsize H}}\!\mathbf{R}_{k} \\
\stackrel{(a)}{=}&\delta\Nb\CG_k \left(\Au_k^* \odot \Ab_k \right) \bigg(\frac{\NoisePower}{\delta\Nb\CG_k}\left(\Au_k^* \odot \Ab_k \right)\herm \Pt^{(\st)}_k \\
&\hspace{5mm} \cdot \left(\Pt^{(\st)}_k\right)\herm \left(\Au_k^* \odot \Ab_k \right)
+ \I   \bigg)^{-1}
 \left(\Au_k^* \odot \Ab_k \right)\herm \:,
\end{align*}
where (a) follows by replacing for $\mathbf{R}_{k}$ from \eqref{eq: R_H} and applying the matrix inversion lemma.
The proof completes by noting that
\be
\Pt^{(\st)}_k \left(\Pt^{(\st)}_k\right)\herm = \frac{\PilotEnergy}{\Np} \left(\Au_k^* \otimes \I \right)
\left(\Au_k^* \otimes \I \right)\herm \:, \nonumber
\ee
and applying Lemma~\ref{lemma: KhatriRao_kronecker}.


\subsection{Proof of Proposition~\ref{prop: bounds-on-estimation-error-S2}}
First, consider that almost surly, $\Au_k$ has linearly independent columns, therefore it is straightforward to show that $\left(\Au_k^* \otimes \I \right)^\dagger \left(\Au_k^* \otimes \I\right) = \I$. Hence we can write
\begin{align*}
\tr\left(\Rt_k^{(\st)}\right) &= \tr\left(\left(\Au_k^* \otimes \I\right)^\dagger \mathbf{Q}_{k}
\left(\left(\Au_k^* \otimes \I\right)\herm\right)^\dagger\right) \\
&\stackrel{(a)}{=} \tr\left(\mathbf{Q}_{k} \left(\mathbf{R}_{\Au_k}\tran \otimes \I\right)^{-1} \right)\; ,
\end{align*}
where $\mathbf{Q}_k = \left(\Au_k^* \otimes \I \right) \Rt_k^{(\st)} \left(\Au_k^* \otimes \I \right)\herm$ and (a) holds due to commutative property of trace.

Now, since both $\mathbf{Q}_k$ and $\left(\mathbf{R}_{\Au_k}\tran \otimes \I\right)^{-1}$ are \ac{PSD} matrices, we can use Lemma~\ref{lemma: bound on trace of  product} and write
\be
\label{eq: bound on sigma_h Sc2}
\frac{\tr\left(\mathbf{Q}_k\right)}{\tr\left(\mathbf{H}_k\right)} \lambda_{\max}^{-1}
\le \sigma_{\Ht_k^{s_2}}^2
\le \frac{\tr\left(\mathbf{Q}_k\right)}{\tr\left(\mathbf{H}_k\right)} \lambda_{\min}^{-1},
\ee
where
\begin{align*}
\lambda_{\max}^{-1} \triangleq \lambda_{\min}\left(\left(\mathbf{R}_{\Au_k}\tran \otimes \I \right)^{-1} \right) &=  \left(\lambda_{\max}\left(\mathbf{R}_{\Au_k} \right)\right)^{-1} \;, \\
\lambda_{\min}^{-1}\triangleq \lambda_{\max}\left(\left(\mathbf{R}_{\Au_k}\tran \otimes \I \right)^{-1} \right) &=  \left(\lambda_{\min}\left(\mathbf{R}_{\Au_k} \right)\right)^{-1} \:.
\end{align*}

Moreover, by taking the same steps as in proof of Proposition~\ref{prop: bounds-on-estimation-error-S1}, it can be shown that
\be
\label{eq: bounds on Q/Rh}
\frac{1}{1\!+\!\delta\Nb\zeta_k}\! \left[1 \!-\! \frac{\epsilon_k^{(\st)}}
{\delta\Nb\zeta_k}\right]^{\!+} \!\!\!\le\! \frac{\tr\left(\mathbf{Q}_{k} \right)}{\tr\left(\mathbf{R}_{k}\right)}\!\le\!
\frac{1}{1\!+\!\delta\Nb\zeta_k} \; .
\ee
By combining the results from \eqref{eq: bounds on Q/Rh} and \eqref{eq: bound on sigma_h Sc2}, the proof is completed.

\subsection{Proof of Corollary~\ref{corollary: pilot_decontamination}}
First note that when $\Nb\to \infty$, $\Ab_k\herm \Ab_j = \mathbf{0}$ for $j\neq k$.
Now, by substituting for $\mathbf{R}_j$ and $\Pt^{(\sth)}_{kj}$ into~\eqref{eq: Q-bar} from~\eqref{eq: R_H} and the line after~\eqref{eq: received signal-s3}, respectively and then using Lemma~\ref{lemma: KhatriRao_kronecker}, it is straightforward to show that $\lim_{\Nb\to\infty} \widebar{\mathbf{Q}}_k\to \boldsymbol{0}$. This implies that the inter-UE interference tends to zero when the number of BS antennas grows large.

In $\sth$, the non-orthogonality of the pilot sequences transmitted from different antennas of a UE leads to intra-UE interference at the BS.
However, $\mathbf{R}_{\Au_k} \to \I_{\Np}$ when $\Nu \to \infty$, implying that the precoded pilots are transmitted through the $\Np$ paths without interfering with each others. Therefore the intra-UE interference tends to zero when the number of UE antennas grows large.

\subsection{Proof of Proposition~\ref{prop: rate in large antenna regime}}
Note that as $\Nb,\Nu \to \infty$, $\Ht^{(\sx)}_k \to \mathbf{0}$ for $\sx = \so,\st$ and $\sth$ according to Proposition~\ref{prop: bounds-on-estimation-error-S1},
\ref{prop: bounds-on-estimation-error-S2} and Corollary~\ref{corollary: pilot_decontamination}, respectively. Therefore, independent from the pilot transmission scenario, in the large antenna regime $\mathbf{R}_{\mathbf{z}_{\text{eff}}} = \NoisePower \I_{\Nb}$.
Substituting for $\mathbf{R}_{\mathbf{z}_{\text{eff}}} = \NoisePower \I_{\Nb}$ and moreover $\mathbf{R}_{\mathbf{x}_k}$ and $\mathbf{F}_k$ from \eqref{eq: covariance of X} and \eqref{eq: data precoder}, respectively, in \eqref{eq: sum-rate} yeilds
\begin{equation}
\begin{split}
    r =& \frac{\Td}{\Tc} \E\left[\log\det\left(\I_{\Nb} + \frac{\DataEnergy}{\Td\Np\NoisePower} \sum_{k=1}^{K} \mathbf{\Sigma}_k \right)\right] \\
    \stackrel{(a)}{\ge}&  \frac{\Td}{\Tc} \log\det\left(\I_{\Nb} + \frac{\DataEnergy}{\Td\Np\NoisePower} \sum_{k=1}^{K} \E\left[\mathbf{\Sigma}_k\right]  \right) \\
=& \frac{\Np \Td}{\Tc} \log\left(1+ \frac{\DataEnergy\sum_{k=1}^{K} \CG_k}{\Td\Np\NoisePower} \right)\: ,
\end{split}
\end{equation}
where (a) is due to Jensen's inequality.

\bibliographystyle{Components/MetaFiles/IEEEtran}
\bibliography{Components/MetaFiles/References}

\end{document}